\documentclass[fleqn,usenatbib]{mnras}

\usepackage{graphicx}	
\usepackage{amsmath}	
\usepackage{amssymb}	
\usepackage{url}
\usepackage{ragged2e}
\usepackage{wrapfig}
\usepackage{textcomp}
\usepackage{hyperref}
\usepackage{setspace}
\usepackage{fancyhdr}
\usepackage{pdfpages}
\usepackage{float}
\usepackage{epstopdf}
\usepackage[toc, page]{appendix}

\usepackage[caption=false]{subfig}

\usepackage{natbib}

\title[Forecasting IDE models using PICO and DESI Mission]{Forecast Analysis on Interacting Dark Energy Models from Future Generation PICO and DESI Missions}

\author[Albin Joseph and Rajib Saha]{
	Albin Joseph\thanks{E-mail: albinje@iiserb.ac.in}
	and Rajib Saha\thanks{E-mail: rajib@iiserb.ac.in}
	\\
$^{1}$Department of Physics, Indian Institute of Science Education and Research (IISER) Bhopal, 462066, India\\
}

\date{Accepted XXX. Received YYY; in original form ZZZ}

\pubyear{2022}

\begin{document}
\label{firstpage}
\pagerange{\pageref{firstpage}--\pageref{lastpage}}
\maketitle

\begin{abstract}
The next-generation CMB satellite missions are expected to provide robust constraints on a wide range of cosmological parameters with unprecedented precision. But these constraints on the parameters could weaken if we do not attribute dark energy to a cosmological constant. The cosmological models involving interaction between dark energy and dark matter can give rise to comparable energy densities at the present epoch, thereby alleviating the so-called cosmic coincidence problem. In the present paper, we perform a forecast analysis to test the ability of the future generation high-sensitive Cosmic Microwave Background (CMB), and Baryon Acoustic Oscillation (BAO) experiments to constrain phenomenological interacting dark energy models. We consider cosmic variance limited future CMB  experiment PICO along with BAO information from the DESI experiment to constrain the parameters of the interacting dark sector. Based on the stability of the cosmological perturbations, we consider two possibilities for the interaction scenario. We investigate the impact of both  coupling constant and  equation of state parameter of dark energy on  CMB temperature power spectrum, matter power spectrum, and $f\sigma_8$. We have used simulated temperature and polarization data from PICO within the multipole ranges ($\ell = 2 - 4000$), and as expected, we do see PICO alone produces better constraints than Planck on the $\Lambda$CDM parameters. With the integration of  PICO and DESI missions, we observe a significant improvement in the constraints on several cosmological parameters, especially the equation of state parameter of dark energy. However, we note that additional data is required to constrain a small positive coupling constant.
\end{abstract}




\section{Introduction}
\label{sec:intro}

The discovery of the accelerated expansion of the universe from the observations of the type 1A supernova~\citep{SupernovaCosmologyProject:1998vns,SupernovaSearchTeam:1998fmf} serves as a breakthrough in the history of cosmology. This discovery gave us the first strong evidence that a large-scale repulsive force could permeate the universe. It also acquainted a new challenge to the cosmologists that the explanation of this phenomenon needs an agent that can cause a repulsive gravity. Moreover, the recent high-precision cosmological observations~\citep{Eisenstein:1998tu,Planck:2018vyg,2010MNRAS.404...60R,BOSS:2016wmc,PhysRevD.103.083533,Joseph:2021omr,Joseph:2021zpn,PhysRevD.98.043526,PhysRevD.98.043528,PhysRevLett.122.171301} also validate the existence of such a repulsive force that steers the accelerated expansion of the universe. The current cosmic acceleration is attributed to an exotic component dubbed 'dark energy', and it provides a major contribution to today's total energy budget of the universe. Recent observations from the Planck satellite collaboration~\citep{Planck:2018vyg} showed that baryonic matter constitutes about $5\%$ of the total energy density of the universe, leaving about $95\%$ of 'dark' components. Among these dark components, the cold dark matter (CDM) accounts for roughly $26\%$ of the energy budget of the universe, whereas the remaining $69\%$ is dominated by dark energy. The dark energy can provide sufficient negative pressure to overcome the gravitational attraction of matter and drives the universe's current accelerating expansion.\\

Among the various cosmological models, the observationally most preferred model is the $\Lambda$CDM  which comprises cold dark matter, baryons, photons, neutrinos, and a cosmological constant $\Lambda$. Despite the successful explanation of  various observed phenomena, the $\Lambda$CDM has been plagued by several theoretical inconsistencies like the coincidence problem~\citep{Steinhardt:2003st,Velten:2014nra} and the cosmological constant problem~\citep{Sahni:1999gb,Sahni:2002kh}. The recent observational data~\citep{Planck:2018vyg} predicts the value of $\Lambda$ to be 123 orders of magnitude smaller than the values estimated from quantum field theory calculations. This huge inconsistency between the theoretical value of $\Lambda$ and its observed value is the cosmological constant problem. The coincidence problem relates to the query of why dark energy and dark matter energy densities are comparable at the present time despite not following the same cosmological evolutionary track. All these inspire cosmologists to search for alternate theories to explain various observed cosmological phenomena. Dark energy is thus introduced as a dynamical entity rather than being thought of as a cosmological constant. Models like holographic dark energy~\citep{Li:2004rb,Zimdahl_2007,PhysRevD.70.043539,Nojiri:2005pu,Zhang_2012,PhysRevD.85.127301,10.1093/mnras/sty903}, Chaplygin gas~\citep{Kamenshchik:2001cp,Bilic:2001cg,PhysRevD.66.043507,PhysRevD.66.081301,PhysRevD.84.123507,PhysRevD.87.083503}, phantom field~\citep{Caldwell:1999ew,PhysRevD.68.023509} and quintom model~\citep{Feng:2004ad,Cai:2006dm} are the other popular choices studied in the literature for the explanation of dark energy.\\

Since the energy densities of dark energy and dark matter are comparable at the present epoch, it opens the possibility that there can be an energy exchange between them. Thus the models with the interaction between dark energy and dark matter are widely studied in the literature~\citep{PhysRevD.61.083503,Pav_n_2004,Amendola_2004,Curbelo_2006,Gonzalez_2006,PhysRevD.76.023508,PhysRevD.77.063513,PhysRevD.78.023505,PhysRevD.78.063527,PhysRevD.78.123514,Quartin_2008,He_2008,PhysRevD.81.043525,PhysRevD.85.103008,PhysRevD.86.103507,PhysRevD.89.083517,PhysRevLett.113.181301,refId0,Caprini_2016,PhysRevD.94.023508,Mukherjee_2017,PhysRevD.99.063517,PhysRevD.103.123547,DiValentino:2020leo,Nunes:2022bhn,Yang:2019vni,Banerjee:2020xcn} as they are very helpful in alleviating this cosmic coincidence problem. Since the models with the interaction between the dark matter and dark energy affect the overall evolution and expansion history of the universe, they are observationally distinguishable from the usual $\Lambda$CDM model. This interaction can then be constrained by the available observational probes like Cosmic Microwave Background (CMB) radiation and Large Scale Structures (LSS) and become a testable theory of the universe. For detailed reviews on the interaction between dark energy and dark matter, we refer to~\cite{Bamba:2012cp},~\cite{Bolotin:2013jpa}, and~\cite{Wang_2016}.\\

Though there are several cosmological observations, they cannot confidently distinguish between the usual $\Lambda$CDM and the interacting dark energy models. Moreover, several interacting scenarios have been confronted with the latest observational data~\citep{PhysRevD.96.123508,Pan:2017ent,PhysRevD.97.043529,Yang_2018,PhysRevD.102.123502}, often a null interaction cannot be discarded with high confidence.  The authors in~\cite{Kumar:2019wfs} and~\cite{PhysRevD.96.103511} demonstrate that the interaction between the dark energy and dark matter can alleviate the tensions on the parameters $H_0$ and $\sigma_8$ simultaneously with excellent accuracy. In~\cite{PhysRevD.95.043520}, new evidence was presented against the $\Lambda$CDM model using data from the Baryon Oscillation Spectroscopic Survey (BOSS) experiment~\citep{refId0boss} of the Sloan Digital Sky Survey (SDSS) collaboration. This evidence was based on measurements of the baryon acoustic oscillations of the Lyman-alpha forest from high redshift quasars. Such deviation from the $\Lambda$CDM can also be explained by the interaction between the dark sectors~\citep{PhysRevD.95.043520}. In this context, the future generation of both astronomical and CMB observational probes will play a significant role in the precise consistency tests of the $\Lambda$CDM model and the other alternate theories, including the interacting dark energy models. In recent years, several efforts have been made to constrain and forecast cosmological parameters in the alternate dark energy scenarios~\citep{Tutusaus:2016kyl,Leung:2016xli,2009JCAP...04..012H}. The forecast analysis in the interacting dark energy models has been performed in ~\cite{Martinelli:2010rt},~\cite{PhysRevD.85.103008},~\cite{PhysRevD.96.103529},~\cite{PhysRevD.84.023504}, and~\cite{Caprini:2016qxs}, however, most of them are outdated since they have explored the forecast of Planck-like CMB surveys or Euclid-like experiments. Moreover, several studies employ the conventional Fisher matrix forecast, which utilizes the information only at the best-fit point and assumes the likelihood function to be Gaussian with respect to the model parameters. But this framework can only provide a reasonable estimate when the likelihood of the parameters approximates a multivariate Gaussian function of the cosmological parameters~\citep{Perotto_2006}. It deviates from the more robust Markov Chain Monte Carlo (MCMC) simulation-based approach especially when the parameters of the model increase~\citep{Perotto_2006}. So in our analysis, we use MCMC exploration of the parameter space as it does not make any assumptions about the Gaussianity or otherwise of the parameter probabilities and is therefore expected to give more reliable results than the traditional Fisher forecast.\\

In this paper, we investigate the ability of future generation high-sensitive CMB experiment~\citep{NASAPICO:2019thw} and baryon acoustic oscillation (BAO) information from the DESI galaxy survey~\citep{2013arXiv1308.0847L} to constrain phenomenological interacting dark energy models. Concerning future CMB observations, we shall consider the cosmic variance limited CMB experiment PICO~\citep{NASAPICO:2019thw}, which is expected to provide high-quality constraints and measure the cosmological parameters with unprecedented precision. We also explore the capability of the CMB mission PICO to constrain the parameters when the dark energy is no longer considered a cosmological constant. For a better insight and to study the effect of dark sector parameters on observables, we perform a theoretical study on the effect of interacting dark energy models on different observable probes like CMB power spectrum, matter power spectrum and $f\sigma_8$. We then exploit the capabilities of complementary observations like BAO information from the DESI galaxy survey to further constrain and provide tighter constrain on the parameters of the interacting dark energy models.\\

We structured our paper in the following way. The section~\ref{Sec:ide_model} is dedicated to the background and perturbation equations of the interacting dark energy models we consider in this work. In section~\ref{Sec:effects_on_probes} we discuss the  effect of interacting dark sector parameters on various observable probes like CMB temperature power spectrum, matter power spectrum and $f\sigma_8$. We then describe the methodology and observational probes for our forecast analysis in section~\ref{Sec:Methods}. After quantifying our results in section~\ref{sec: results} through 2D posterior distributions and mean values with $68\%$ and $95\%$ confidence levels, we discuss and draw our conclusions in the last section~\ref{sec: conclusion}.

\section{Interacting Dark Energy Models} \label{Sec:ide_model}

The space-time interval $ds$ in the spatially flat homogeneous and isotropic model of the universe is described by,
\begin{align}\label{eq: metric}
	ds^2 = a^2(\tau)(-d\tau^2 + \delta_{ij}dx^idy^j),
\end{align}
where $\tau$ is the conformal time and is related to the comic time $t$ as $a^2d\tau^2 = dt^2$. Moreover $a(\tau)$ is the scale factor and it satisfies the Friedmann equation,
\begin{equation}\label{eq: hubble}
	\small
	\mathcal H^2 \equiv \left(\frac{a^{'}}{a}\right)^2 = \frac{8\pi G}{3} a^2 \rho_{tot} = \frac{8\pi G}{3} a^2 \left(\rho_{\gamma} + \rho_{\nu}+\rho_b + \rho_c +\rho_{x}\right).
	\small
\end{equation}
Here $\rho_{tot}$  represents the total background energy density of all different components of the universe. They include neutrinos ($\nu$), photons ($\gamma$), cold dark matter ($c$), baryons ($b$),  and dark energy ($x$). A prime indicates the differentiation with respect to the conformal time. In this article, we assume that the energy transfer is only between dark energy and dark matter such that their conservation equation reads,
\begin{align}
	\label{eq: cons_eqs 1} \rho'_{c}+3 \mathcal{H} \rho_{c} &=Q\,, \\ \rho'_{x}+3 \mathcal{H}(1+w_x) \label{eq: cons_eqs 2} \rho_{x} &=-Q\,.
\end{align}
The cold dark matter is assumed to be pressureless i.e., $p_c =0$. Since the other three fluids-- neutrinos ($\nu$), photons ($\gamma$), and baryons ($b$) conserve independently, they have no energy transfer between them. Thus their conservation equations can be written as,
\begin{align}
	\rho'_{A}+3 \mathcal{H}(1 + \omega_A) \rho_{A} &=0\,,
\end{align}
here $\omega_A= p_A/\rho_A$ represents the equation of state parameter of the A$^{th}$ fluid where A = $\gamma, \nu, b$. For the neutrinos and photons, $\omega_{\nu} = \omega_{\gamma} =1/3$, whereas for  cold dark matter and baryons $\omega_c = \omega_b =0$. The equation of state parameter of the dark energy is given by, $\omega_x = p_x/\rho_x$. Moreover, $Q$ in eq.~\ref{eq: cons_eqs 1} and eq.~\ref{eq: cons_eqs 2} encodes the interaction between the dark sector. The energy transfer is from dark matter to dark energy if the interaction function $Q < 0$, whereas for $Q >0$, the energy exchange is from dark energy to dark matter. For $Q =0$, one can obtain the usual $\Lambda$CDM model. Using eq.~\ref{eq: cons_eqs 1} and ~\ref{eq: cons_eqs 2}, the effective background equation of state for the dark energy and dark matter can be written as,
\begin{equation}\label{eq: eff_c}
	w_{c}^{\rm eff} =-\frac{Q}{3 \mathcal{H} \rho_{c}},
\end{equation}
\begin{equation}\label{eq: eff_x}
	w_{x}^{\rm eff} =w_x+\frac{Q}{3 \mathcal{H} \rho_{x}}.
\end{equation}
Here $\omega_x$ is the equation of state of dark energy for a vanishing interaction. The eq.~\ref{eq: eff_c} and eq.~\ref{eq: eff_x} show how the interaction between the dark sector can affect the effective dark energy equation of state parameter. Moreover, they show that positive (negative) values of $Q$ contribute to an effective negative (positive) pressure in the background equation. As a result, the dark matter content in the past will be less (more) compared to the uncoupled case, and as a consequence, the epoch of the radiation-matter equality will be delayed (earlier). This decrease (increase) in the amount of matter content of the universe has a significant effect on the CMB power spectrum and matter power spectrum, which we further investigate in section~\ref{Sec:effects_on_probes}. 
\\
\begin{figure*}
	\centering
	\subfloat{\includegraphics[width=.5\linewidth]{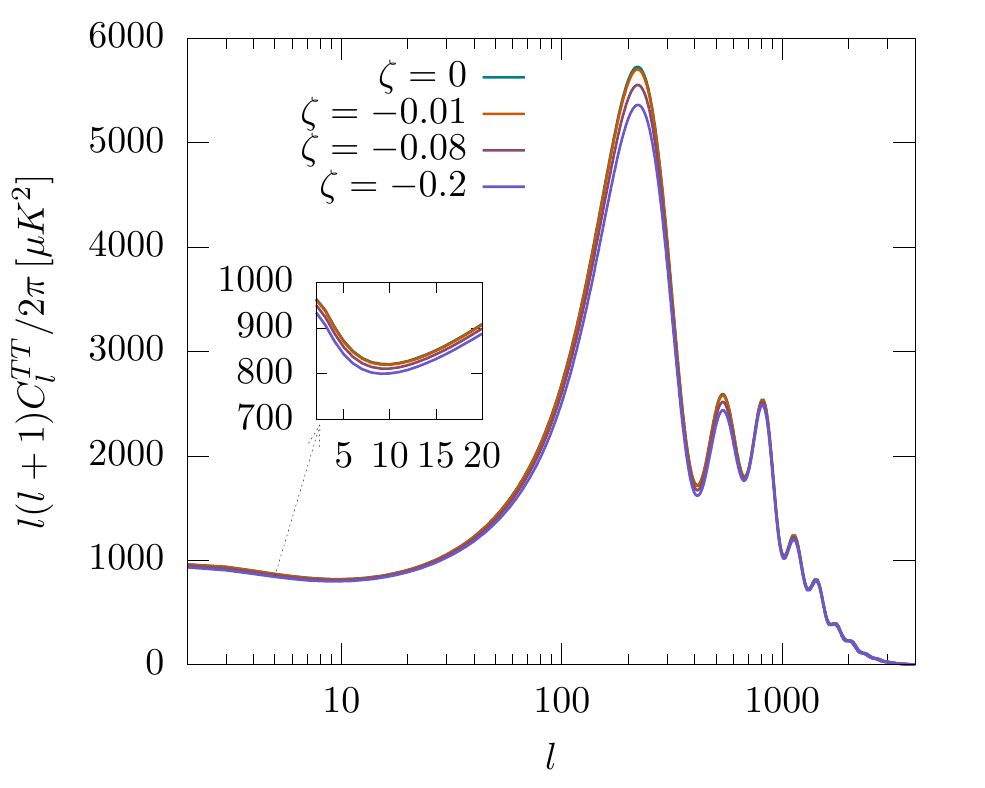}}\hfill
	\subfloat{\includegraphics[width=.5\linewidth]{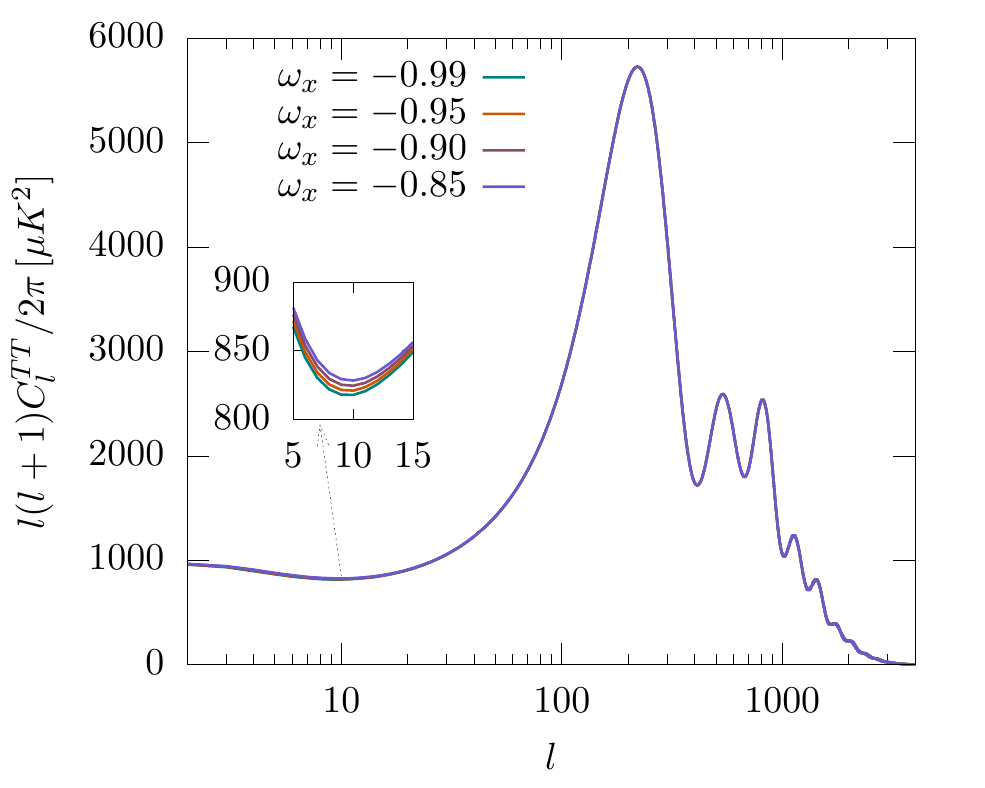}}\hfill
	\subfloat{\includegraphics[width=.5\linewidth]{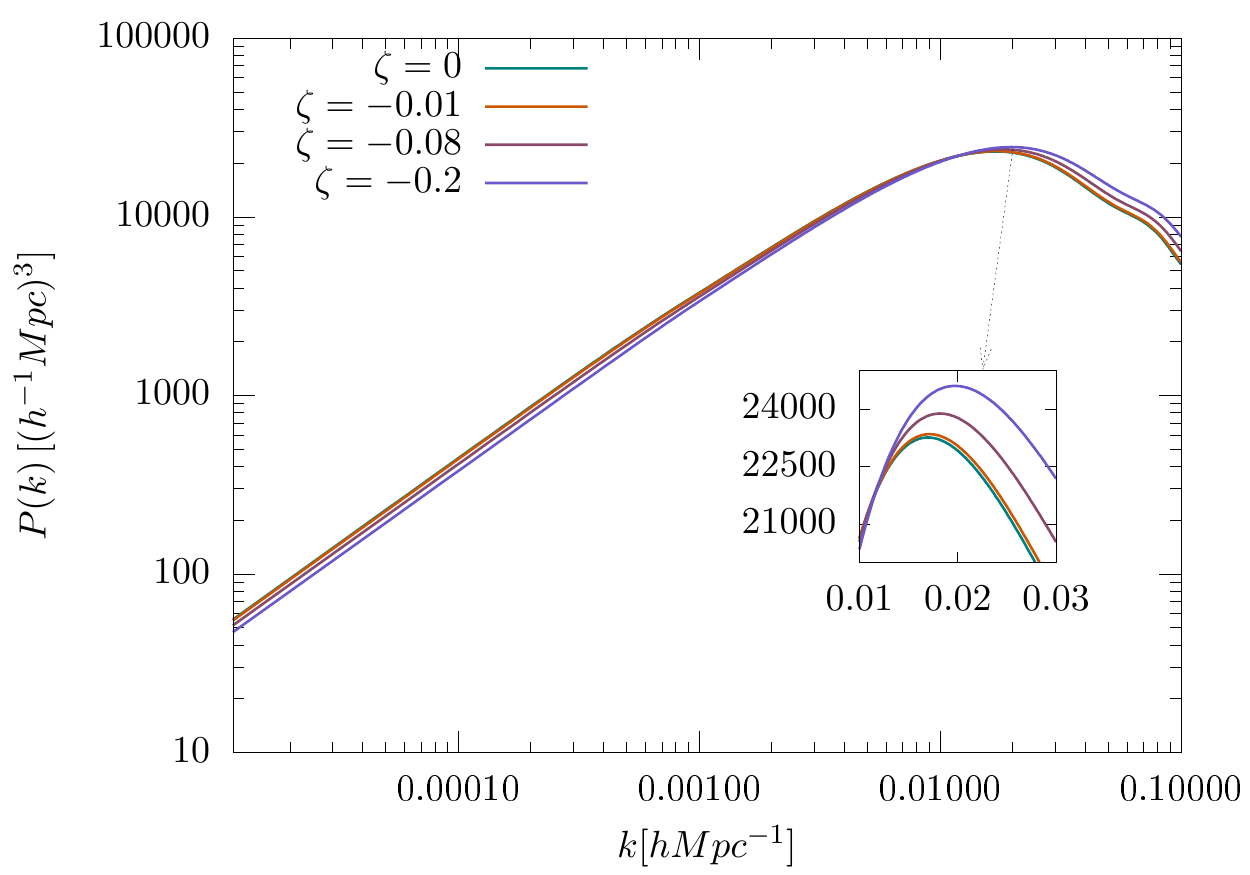}}\hfill
	\subfloat{\includegraphics[width=.5\linewidth]{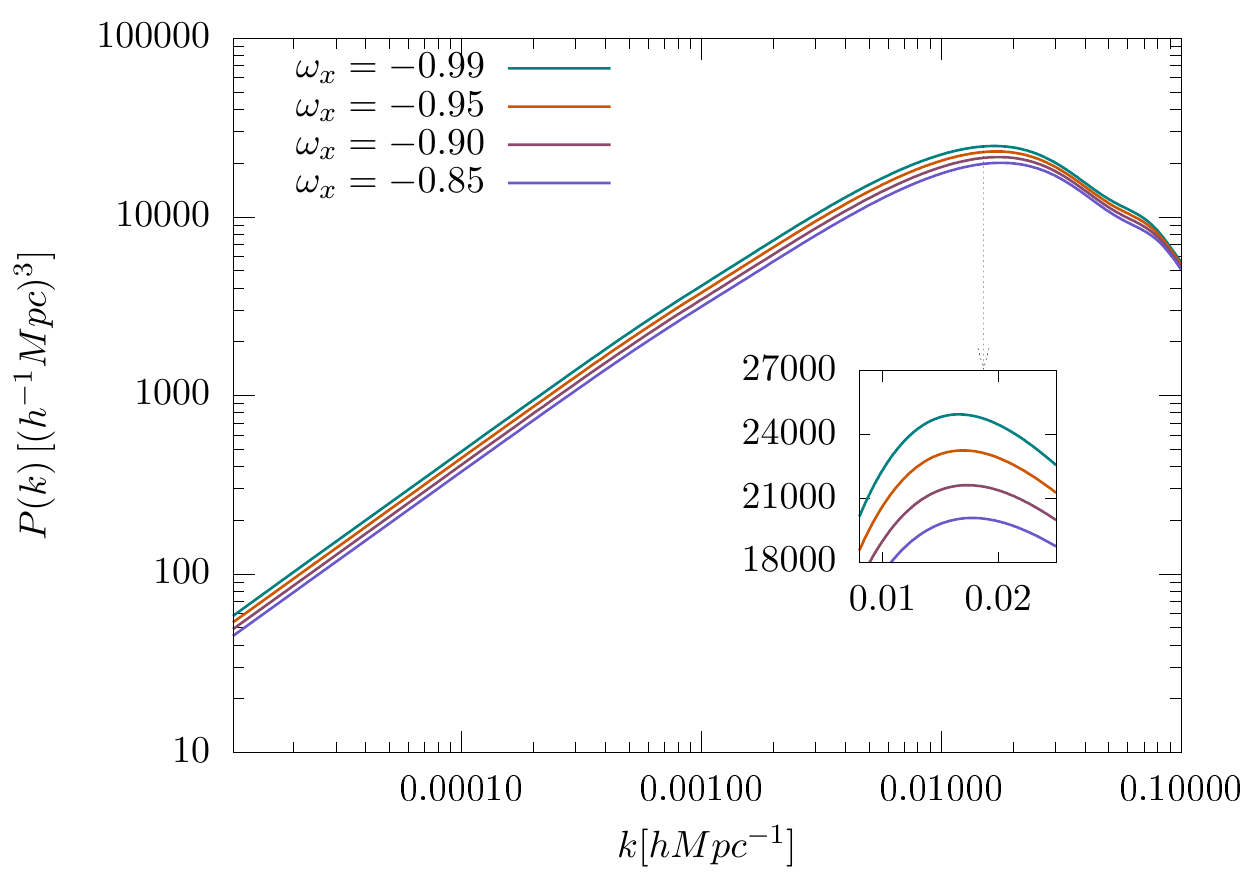}}\hfill          
	\caption{The left panel shows the effect of coupling constant $\zeta$ on CMB temperature power spectrum and matter power spectrum for model 1 with $\omega_x=-0.95$. We utilise the fiducial values listed in the table~\ref{tab:model1_table} for the remaining parameters. As the coupling becomes more negative, the CMB acoustic peak positions are further suppressed. The right panel shows the effect of dark energy equation of state parameter $\omega_x$ on CMB and matter power spectrum for the same model with $\zeta=-0.08$. The variations in the $\omega_x$ mainly affect the low-$\ell$ modes of the CMB angular power spectrum and can also cause an overall shift in the amplitude of the matter power spectrum.}\label{im:ps_model1}
\end{figure*}
In the case of energy transfer between dark energy and dark matter, the energy densities of the dark sector evolve differently for different models. There are different forms of phenomenological interaction functions proposed in the literature. Among them the popular choices are the interacting models with $Q$ proportional to either $\rho_x$ or $\rho_c$ or any combination of them~\citep{PhysRevD.78.023505,PhysRevD.89.083517,PhysRevD.97.043529,PhysRevD.85.043007,PhysRevD.89.103531}. In this work, we investigate a very well-studied phenomenological energy transfer function between dark energy and dark matter, which can be expressed as,
\begin{align}\label{eq: Q}
	Q=\zeta \mathcal{H} \rho_x \,,
\end{align}
where $\zeta$ is the coupling constant and $\rho_x$ is the energy density of the dark energy. For $\zeta=0$ we recover the uncoupled case of standard cosmology. In~\cite{V_liviita_2008},~\cite{He:2008si}, and~\cite{10.1111/j.1365-2966.2009.16140.x} it has been shown that due to the presence of interaction term in the non-adiabatic perturbations of the dark energy the gravitational instabilities arise for a constant $\omega_x \simeq -1$. Moreover, the early time instabilities in the evolution of the dark energy perturbations~\citep{PhysRevD.85.043007,V_liviita_2008,He:2008si,Gavela_2009,PhysRevD.79.043526,PhysRevD.79.063518,PhysRevD.79.043522,PhysRevD.80.103514,Gavela_2010} depend on the the parameters $\zeta$ and $(1 + \omega_x)$ via a ratio called the dooms factor, given as
\begin{align}\label{eq: dooms}
	d \equiv \frac{Q}{3 \mathcal{H} \rho_x (1 + \omega_x)}.
\end{align}
One can avoid the early time instabilities if the dooms factor $d$ is negative semidefinite ($d\leq 0$)~\citep{Gavela_2009}, which ensures the parameters $\zeta$ and ($1 + \omega_x$) to be of opposite signs. So stable perturbations can be achieved by considering energy exchange between dark matter to dark energy ($\zeta <0$) with non phantom or quintessence equation of state (EoS) $(1 + \omega_x >0 )$ or energy exchange from dark energy to dark matter ($\zeta > 0$) with phantom EoS $(1 + \omega_x < 0)$. Thus in this work, based on the stability of cosmological perturbations, we consider two cases of interacting scenarios, given by,
\begin{align}\label{eq: models}
	\zeta < 0 ,\,\, -1 < \omega_x < -1/3 : \,\,\,\text{MODEL\, 1},\\
	\nonumber
	\\ 
	\zeta > 0 ,\,\, \omega_x < -1 : \,\,\,\text{MODEL\, 2}.
\end{align}
The energy densities of dark energy and dark matter can be obtained analytically by solving eq.~\ref{eq: cons_eqs 1} and ~\ref{eq: cons_eqs 2}. For our choice of interacting function, Q in eq.~\ref{eq: Q}, the energy densities of the dark sector for model 1 and model 2 are given by,
%
\begin{align}
	& \rho_{c}=\rho_{c, 0} a^{-3}+\zeta\frac{\rho_{x, 0}a^{-3}}{3w_{x}^{\rm eff}}\left[1-a^{-3w_{x}^{\rm eff}}\right]\,, \\
	& \rho_{x}=\rho_{x, 0} a^{-3(1+w_{x}^{\rm eff})}\,,
\end{align}
where $w_{x}^{\rm eff}$ is the effective equation of state for the dark energy given in eq.~\ref{eq: eff_x}. Apart from these background modifications to the $\Lambda$CDM model, the other modifications are at the perturbation level. In the synchronous gauge, we get~\citep{Gavela_2010,Honorez_2010,PhysRevD.96.043503,DiValentino:2019ffd,PhysRevD.101.063502,PhysRevD.88.023531}
\vspace{-0.11cm}
\begin{align}
	\label{eq: pert_1} \delta_{c}^\prime= &-\theta_{c}-\frac{h^\prime}{2}\left(1-\frac{\zeta}{3} \frac{\rho_{x}}{\rho_{c}}\right) +\zeta \mathcal{H} \frac{\rho_{x}}{\rho_{c}}\left(\delta_{x}-\delta_{c}\right)\,, \\ 
	\label{eq: pert_3} \theta_{c}^\prime=&-\mathcal{H} \theta_{c}\,, 
	\\ \nonumber \delta_{x}^\prime=&-(1+w_x)\left[\theta_{x}+\frac{h^\prime}{2}\left(1+\frac{\zeta}{3(1+w_x)}\right)\right]+ \\ & \hspace{0 cm} -3 \mathcal{H}(1-w_x)\left[\delta_{x}+\frac{\mathcal{H} \theta_{x}}{k^{2}}(3(1+w_x)+\zeta)\right]\,, \\ 
	\label{eq: pert_2} \theta_{x}^\prime=& 2 \mathcal{H} \theta_{x}\left[1+\frac{\zeta}{1+w_x} \left(1-\frac{\theta_{c}}{2\theta_{x}}\right)\right]+\frac{k^{2}}{1+w_x} \delta_{x}\,,
\end{align}
where $\delta$ and $\theta$ denote the density contrasts and velocities of the dark sector. Moreover, $h$ is the synchronous gauge field defined in the Fourier space, and $k$ is the wave number. The initial conditions for the dark energy perturbations are given by~\citep{PhysRevD.96.043503,PhysRevD.88.023531},
\begin{align}
	\delta_x^{\rm in}(\eta) &=& \frac{3}{2}\frac{(2\zeta-1-w_x)(1+w_x+\zeta/3)}{12w_x^2-2w_x-3w_x\zeta+7\zeta-14}\delta_{\gamma}^{\rm in}(\eta)\,,\\
	\theta_x^{\rm in} &=& \frac{3}{2}\frac{k\eta(1+w_x+\zeta/3)}{2w_x+3w_x\zeta+14-12w_x^2-7\zeta}\delta_{\gamma}^{\rm in}(\eta)\,,
	\label{eq:initialconditions}
\end{align}
where $\eta = k\tau$ and $\delta_\gamma^{in}(\eta)$ denotes the initial condition for the photon density perturbations. Moreover, the dark energy sound speed has fixed to unity, i.e., $c_{s,x}^2 = 1$ and the dark energy adiabatic sound speed is given by $c_{a,x}^2 = w_x$ (for more details we refer to the work in~\cite{V_liviita_2008}).

\section{Effect on Observable Probes} \label{Sec:effects_on_probes}
\vspace{-0.15cm}
In this section, we investigate the impact of the interacting dark energy models (models 1 and 2) on several observable quantities, such as the $f\sigma_8$, the CMB temperature power spectrum, and the matter power spectrum. For this analysis, to compute the power spectrum, we implement the background and perturbation equations (see section~\ref{Sec:ide_model}) in the publicly available Boltzmann code CLASS (V$3.2.0$)~\citep{Blas_2011,2011arXiv1104.2932L}.  In our current analysis, we focus on the study of matter power spectrum in the linear regime\footnote{ To model the non-linear regime of the matter power spectrum, we refer to the works in~\cite{PhysRevD.105.123506} and~\cite{PhysRevD.102.063533}.} where $k\leq0.1\,\, h\,Mpc^{-1}$. Furthermore, for the cosmological parameters (excluding the values of coupling constant $\zeta$ and equation of state parameter $\omega_x$) we adopt the fiducial values given in table~\ref{tab:model1_table} and table~\ref{tab:model2_table} for model 1 and model 2 respectively.

\subsection{CMB Sector} \label{Sec:cmb_sector}
\begin{figure*}
	\centering
	\subfloat{\includegraphics[width=.5\linewidth]{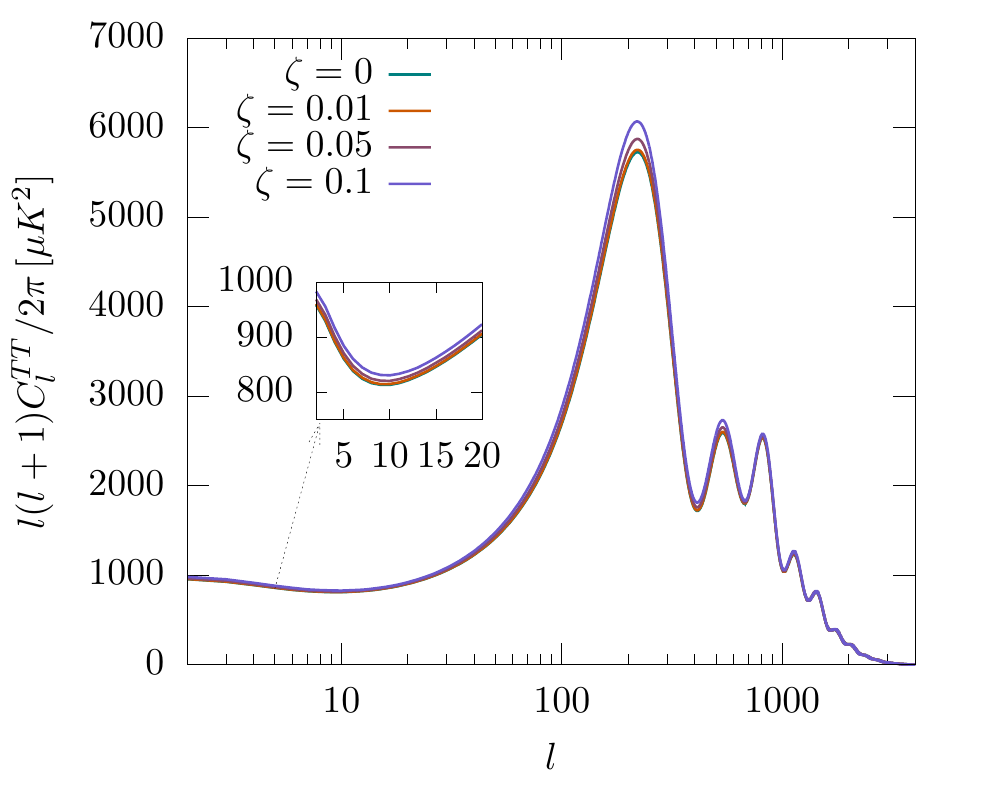}}\hfill
	\subfloat{\includegraphics[width=.5\linewidth]{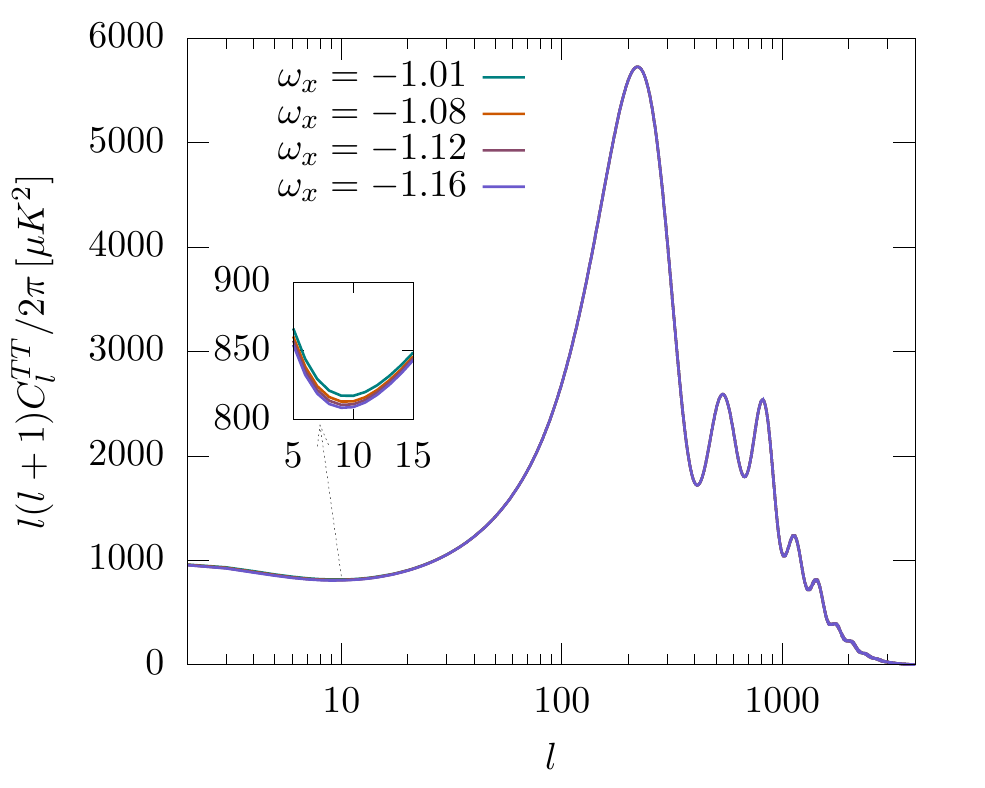}}\hfill
	\subfloat{\includegraphics[width=.5\linewidth]{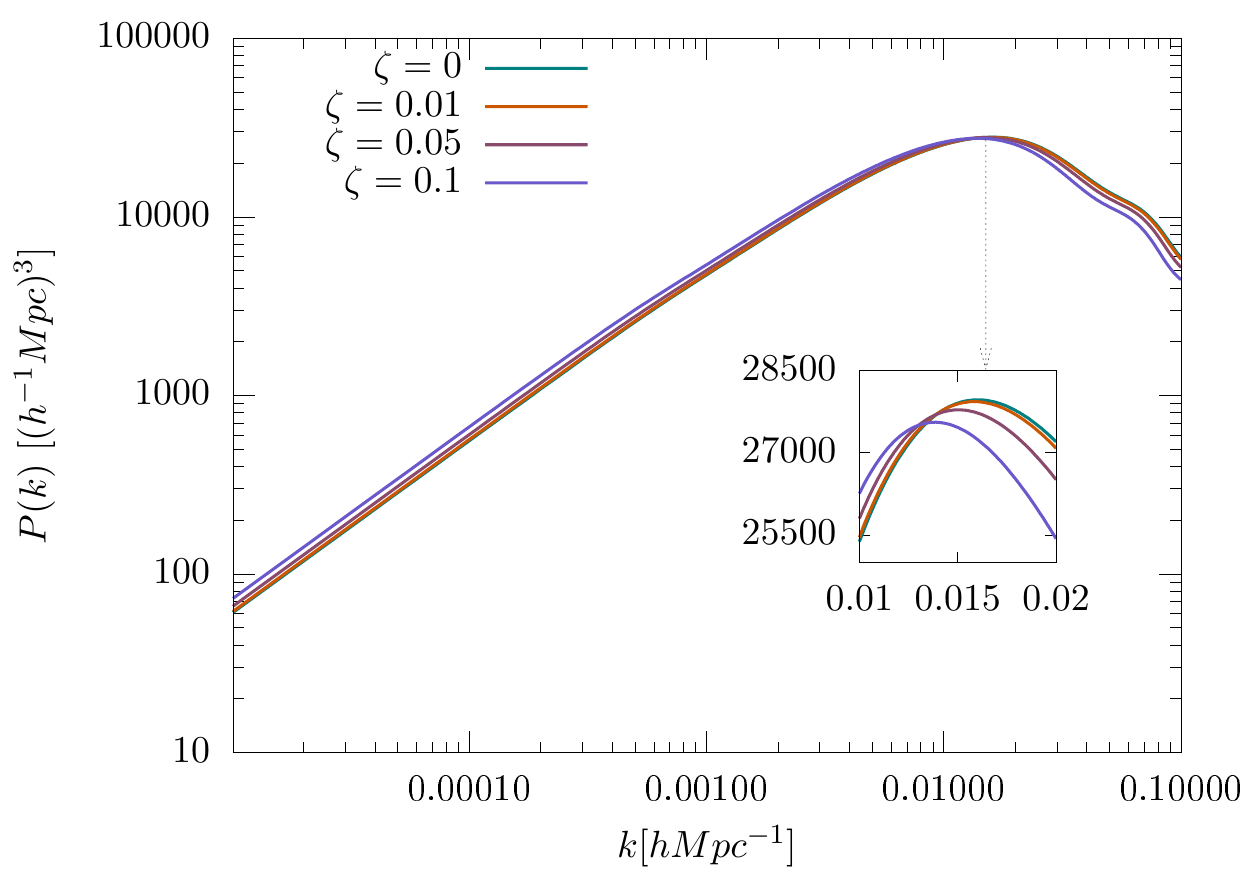}}\hfill
	\subfloat{\includegraphics[width=.5\linewidth]{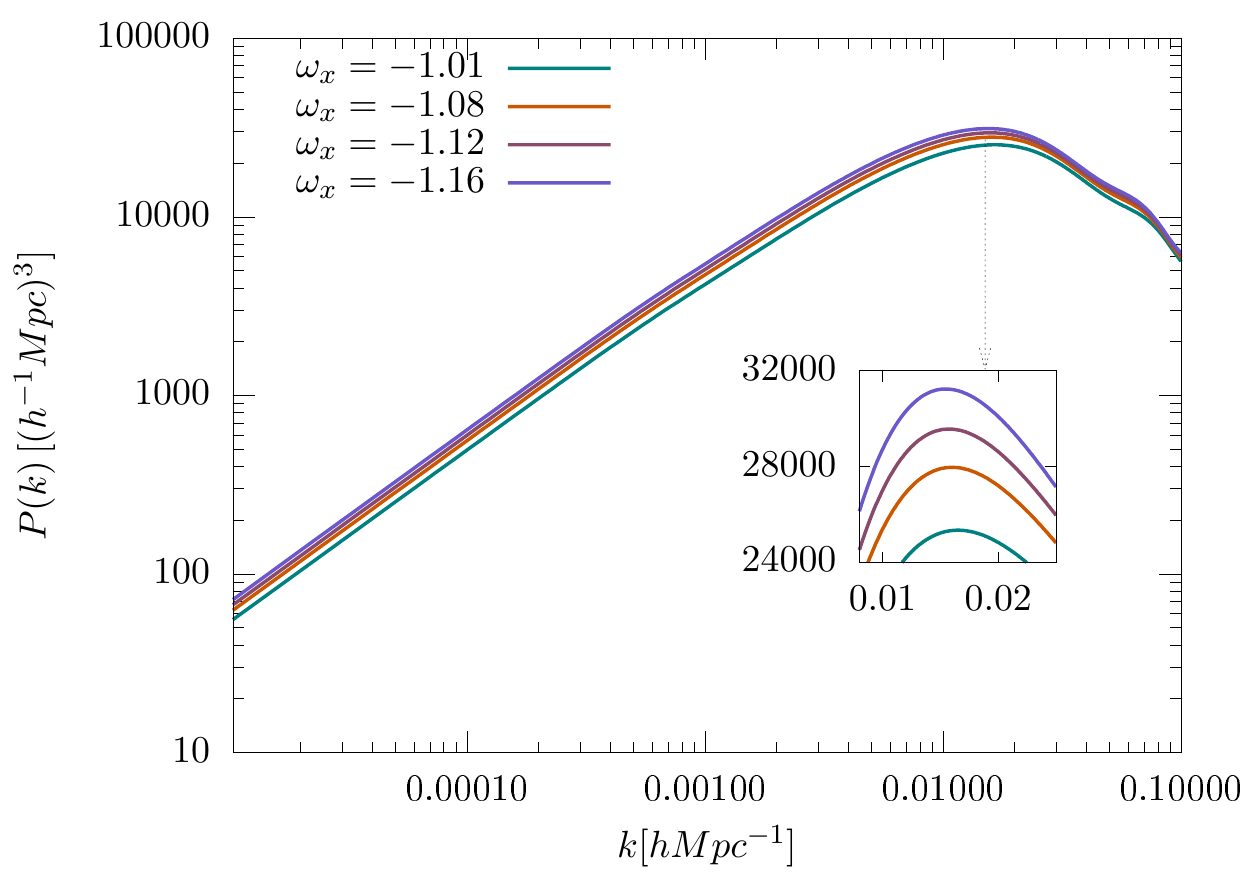}}\hfill          
	\caption{The left panel shows the effect of coupling constant $\zeta$ on CMB temperature power spectrum and matter power spectrum for model 2 with $\omega_x=-1.08$. We utilise the fiducial values listed in the table~\ref{tab:model2_table} for the remaining parameters. As the coupling becomes more positive, the amplitude of the matter power spectrum is further suppressed. The right panel shows the effect of the dark energy equation of state parameter $\omega_x$ on CMB and matter power spectrum for the same model with $\zeta=0.05$. The variations in the $\omega_x$ only affect the low-$\ell$ modes of the CMB angular power spectrum while leaving the acoustic peaks almost untouched.}\label{im:ps_model2}
\end{figure*}

The CMB radiation is the relic radiation from the very early stages of the universe that has a significant role in understanding the evolutionary history of the universe. The CMB temperature power spectrum is given as,
\begin{equation}
	C_\ell^{TT} = \frac{2}{k}\int k^2 dk P_\xi (k) \Delta_{T\ell}^2(k)
\end{equation}
where $P_\xi (k)$ is the primordial power spectrum, $\ell$ is the multipole index and $\Delta_{Tl}^2(k)$ represents the temperature transfer function. We refer to the studies in~\cite{1995ApJ...444..489H} and~\cite{1996ApJ...469..437S} for a detailed analysis of the CMB power spectrum. The numerical evaluation of the CMB temperature power spectrum for the interacting dark energy scenarios is implemented in the latest version of the cosmological Boltzmann integrator code CLASS (V3.2.0)~\citep{Blas_2011,lesgourgues2011cosmic}. The theoretical computation results of the CMB power spectrum for various values of the coupling constant $\zeta$ and the dark energy equation of state parameter $\omega_x$ are shown in the top panel of figure~\ref{im:ps_model1} and ~\ref{im:ps_model2}.

The time of transition from a radiation-dominated epoch to a matter-dominated epoch of the universe is greatly influenced by the amount of dark matter content in the universe. A lower (higher) dark matter density in the universe can cause a delay (early) in this transition epoch. As a result, the universe enters matter dominated later (earlier) than the case for the universe with more (less) matter content. This leads to a scenario where the universe becomes more (less) radiation dominated in the early phases of its evolution, which further affects the gravitational potential wells. This phenomenon which occurs in the very early universe, causes an increase (decrease) in the so-called `radiation driving effects'~\citep{Hu_2002}, which manifests as a rise (lower) in the acoustic peak positions of the CMB power spectrum. In the top left of figure~\ref{im:ps_model1} we show the effect of the coupling parameter $\zeta$ on the CMB temperature power spectrum for the case of model 1. Apart from $\zeta$, the rest of the cosmological parameters are assumed to take the fiducial values given in table~\ref{tab:model1_table}. For model 1, as the coupling constant $\zeta$ is negative, the dark matter density is higher in the past compared to the $\Lambda$CDM model. As a result, the transition epoch will be earlier, which ultimately manifests as an overall suppression in the CMB acoustic peak position (see top left of figure~\ref{im:ps_model1}). As the coupling becomes more negative, the peak positions are further suppressed. Moreover, the presence of interaction also lowers the low-$\ell$ CMB power spectrum through the Integrated Sachs-Wolfe (ISW) effect (see inset in the top left of figure~\ref{im:ps_model1}). The variation of the constant equation of state parameter of dark energy $\omega_x$ (for a fixed $\zeta = -0.08$ for model 1) only modifies the low-$\ell$ part of the CMB power spectrum, and it leaves the acoustic peaks almost untouched (see top right of figure~\ref{im:ps_model1} and its inset). 

In the top left of figure~\ref{im:ps_model2}, we show the effect of the coupling parameter $\zeta$ on the CMB temperature power spectrum for the case of model 2. In this case, the cosmological parameters apart from $\zeta$ are assumed to take fiducial values given in table~\ref{tab:model2_table}. Since the coupling constant $\zeta$ is positive for model 2, there is less dark matter in the past compared to the $\Lambda$CDM. In this case, the dark energy gets converted to dark matter as time evolves. A lower dark matter density in the past corresponds to a delay in the transition epoch, and due to the radiation driving effects, there is a rise in the peak positions of the CMB power spectrum. As the coupling becomes more positive, the acoustic peak positions are further enhanced (see top left of figure~\ref{im:ps_model2}). Moreover, the coupling also enhances the low-$\ell$ modes of the CMB power spectrum through the Integrated Sachs-Wolfe (ISW) effect (see inset in the top left of figure~\ref{im:ps_model2}). It is also worth noting that the variation of the constant $\omega_x$ (for a fixed $\zeta = 0.05$ for model 2) only modifies the low-$\ell$ part of the CMB power spectrum and leaves the acoustic peaks almost untouched (see top right of figure~\ref{im:ps_model2} and its inset). Since large couplings can cause a more pronounced effect on the acoustic peaks (see top left of figure~\ref{im:ps_model1} with $\zeta=-0.2$ and figure~\ref{im:ps_model2} with $\zeta=0.1$), such large couplings can be easily excluded by the observational data. Moreover, the small couplings can cause very subtle changes that are hard to distinguish, and from the previous analysis, small couplings are preferred by the observational data~\citep{PhysRevD.95.043520,PhysRevD.83.063515, Costa_2017, Yang_2016}. Thus for our analysis, we choose the fiducial values of the coupling constant $\zeta$ and the equation of state parameter of dark energy $\omega_x$ to be within $1\sigma$ of the best-fit given in~\citep{Costa_2017} i.e., ($-0.08, -0.95$) and ($0.05, -1.08$) for model 1 and model 2 respectively. For the rest of the cosmological parameters, we use the parameter values from the recent Planck 2018 results~\citep{Planck:2018vyg}.

\subsection{Matter Sector} \label{Sec:matter_sector}
A powerful measure of the statistical distribution of matter is the matter power spectrum, and it plays a significant role in comprehending the dynamics of our universe. The observed large-scale structures are the consequence of the amplification of primordial density fluctuations by gravitational instability. The matter power spectrum~\citep{dodelson} is given by,
\begin{equation} \label{power}
	P\left(k,a\right)= A_s \,k^{n_s} T^2\left(k\right) D^2\left(a\right),
\end{equation}
where $n_s$ is the spectral index, $D(a)=\frac{\delta_m(a)}{\delta_m(a=1)}$ represents the normalized density contrast, $A_s$ is the scalar primordial power spectrum amplitude, and $T(k)$ is the matter transfer function. The combined effects of the complementary actions of the inward pull by gravity and the outward push by the radiation pressure, which are accountable for the CMB acoustic oscillations, also significantly affect the matter power spectrum.    As we discussed in the previous section, due to the energy flow from dark matter to dark energy in model 1, the dark matter density is higher in the past compared to the $\Lambda$CDM model. As the coupling constant $\zeta$ becomes more negative, there is an increase in the dark matter content of the universe, which in turn causes a rise in the matter power spectrum (see bottom left of figure~\ref{im:ps_model1} and its inset). The changes in the dark energy equation of state parameter $\omega_x$, which mainly affects the low-$\ell$ modes of the CMB angular power spectrum, can also cause an overall shift in the amplitude of matter power spectrum (see bottom right of figure~\ref{im:ps_model1} and its inset). Moreover, from figure~\ref{im:ps_model1}, it is interesting to note that the effect of the variation of the dark energy equation state parameter is more pronounced for the case of matter power spectrum than CMB power spectrum.
\begin{table}
	\addtolength{\tabcolsep}{8pt}
	\renewcommand{\arraystretch}{1.4}
	\centering
	\begin{tabular}{lc} 
		\hline 
		\hline
		Model   &   $\sigma_8$ \\ 
		\hline 
		\hline
		Model 1
		&  $0.8805$
		\\
		$\Lambda$CDM
		& $0.8224$
		\\
		Model 2
		& $0.78819$
		\\
		\hline 
		\hline
	\end{tabular} 
	\caption{\small
		The values of $\sigma_8$ at redshift $z=0$ for model 1, model 2 and the $\Lambda$CDM. \normalsize
		\label{tab:sigma8}
	}
\end{table}

In the bottom panel of figure~\ref{im:ps_model2}, we show the effect of $\zeta$ and $\omega_x$ on the matter power spectrum for the case of model 2. Since $\zeta$ is positive for model 2, the energy flow is from dark energy to dark matter. As a result, the dark matter content in the past will be less compared to the $\Lambda$CDM model. A decrease in the matter content of the universe can cause a decrease in the matter power spectrum. As the coupling becomes more positive, the amplitude of the matter power spectrum is further suppressed (see bottom left of figure~\ref{im:ps_model2} and its inset). Any dynamical change that decreases the matter power spectrum's amplitude correlates to a decline in the Newtonian potential that increases the degree of anisotropy~\citep{Hu_2002}. Consequently, a decline in the matter content
due to the presence of a positive interaction constant of model 2 drives the CMB spectrum higher and the matter power spectrum down. Moreover, though the variation of $\omega_x$ for a fixed $\zeta$ leaves the CMB acoustic peaks untouched, it has a significant impact on the amplitude of the matter power spectrum. One can see a more pronounced shift in the amplitude of the matter power spectrum as the dark energy equation of state parameter deviates from $\omega_x = -1.0$ (see bottom right of figure~\ref{im:ps_model2} and its inset). 

\begin{figure}
	\centering
	\includegraphics[width=0.49\textwidth]{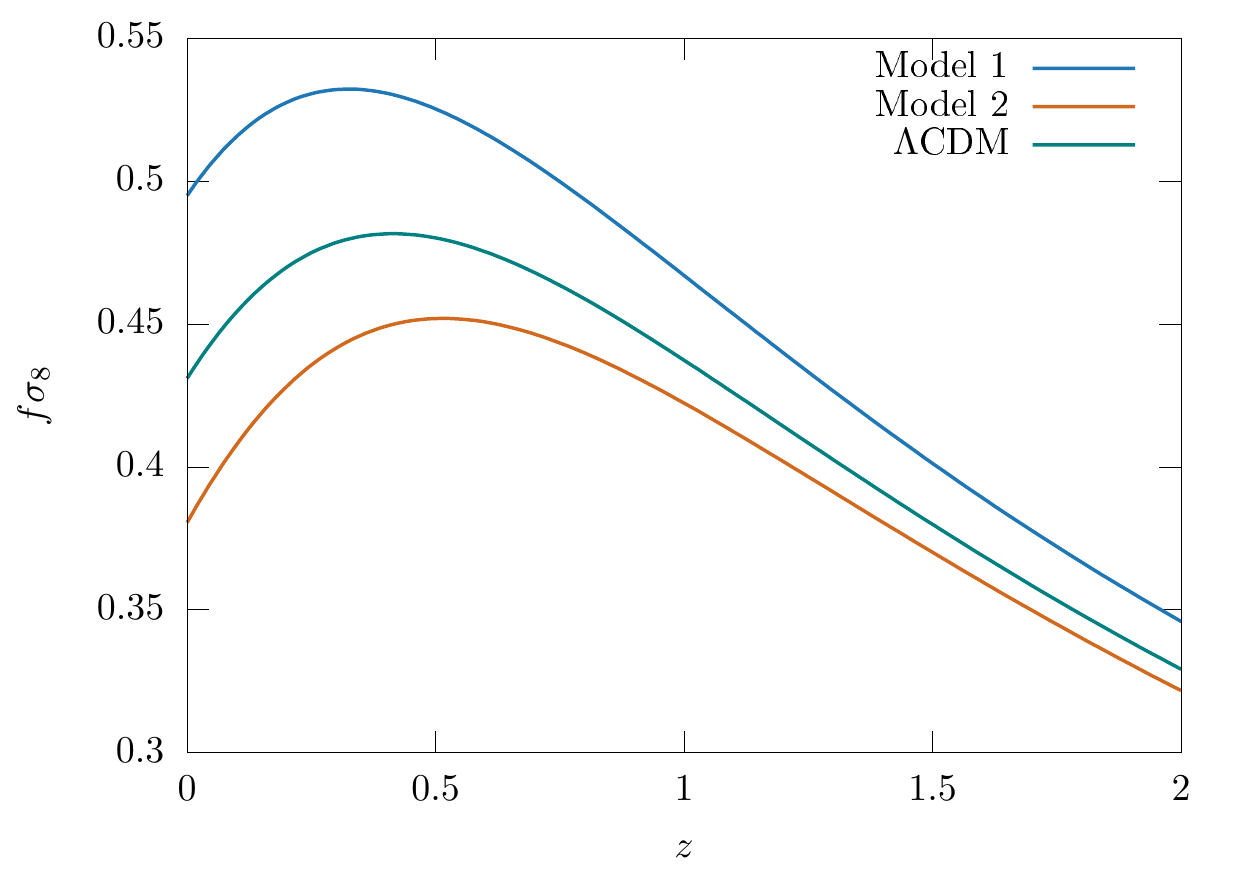}
	\caption{\small Figure depicting the variation of $f\sigma_8$ in the low redshift zone as a function of redshift $z$. At the present epoch $z=0$, the $f \sigma_8$ of model 1 is the highest compared to the $\Lambda$CDM and model 2. \normalsize}
	\label{fig:sigma}
\end{figure}

The linear growth rate is yet another excellent tool for distinguishing between various dark energy theories based on the growth of large-scale structures. The linear growth rate can be defined as,
\begin{equation} \label{eq:growth_rate}
	f(a)= \frac{d \ln \delta_{m}}{d \ln a}~=~ \frac{a}{\delta_{m}(a)}\frac{d \delta_{m}}{d \,a}~.
\end{equation}
Due to the presence of interaction between the dark sectors, the linear growth rate gets modified, and it further helps to differentiate dark energy models even better. Another more reliable and powerful observational quantity which is measured by the redshift surveys is the product of $f(a)\sigma_8$~\citep{Percival_2009}, where $\sigma_8$ is defined as the root-mean-square (rms) fluctuations of the linear density field within the sphere of radius $R = 8 \,h^{-1}$Mpc. In the linear regime, $\sigma_8$ and $f \sigma_8$ can be written as~\citep{PhysRevD.77.023504, Song_2009,Huterer_2015,Ishak_2018},
\begin{equation}
	\sigma_8(z) = \sigma_8(z=0)\frac{\delta_{m}(z)}{\delta_{m}(z=0)},
\end{equation}

and
\begin{equation} \label{eq:fsigma}
	f \sigma_8(z) \equiv f(z)\sigma_8(z) =-(1+z)\frac{\sigma_8(z=0)}{\delta_{m}(z=0)}\frac{d \delta_{m}}{d \,z}~.
\end{equation}
\begin{figure}
	\centering
	\includegraphics[width=0.49\textwidth]{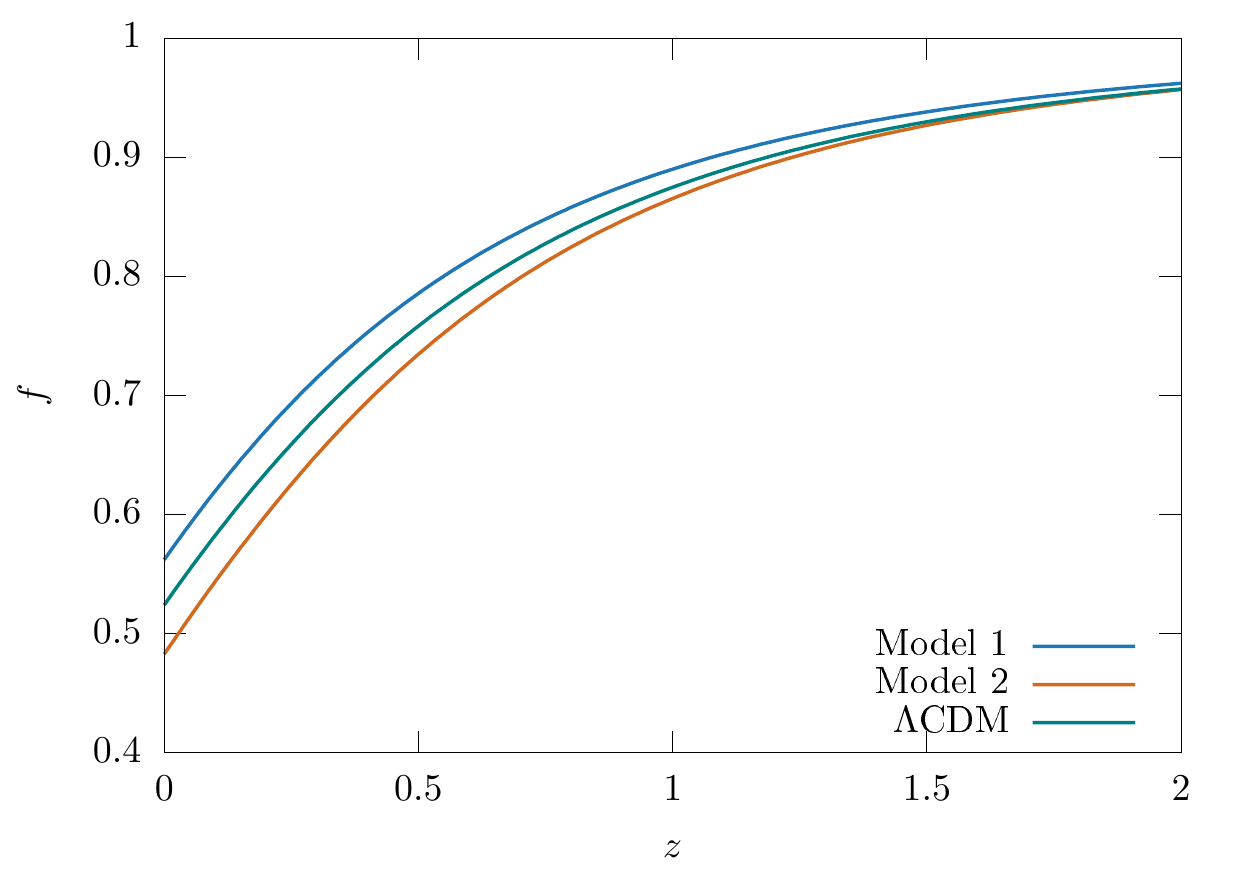}
	\caption{\small  Figure depicting the variation of linear growth rate $f$ in the low redshift zone as a function of redshift $z$. The linear growth rate $f$ for model 1 is slightly higher compared to the $\Lambda$CDM, whereas $f$ is lowest for the case of model 2. \normalsize}
	\label{fig:growth}
\end{figure}
The $\sigma_8(z=0)$ represents the value of the rms fluctuations of the linear density field at $z=0$. The scale factor $a$ is related to the redshift $z$ by the relation $z=\frac{a_0}{a}\,-1$ where $a_0$ represents the present value of the scale factor. Since the $f\sigma_8$ is a more dependable quantity~\citep{Percival_2009}, it provides a better comprehension of the evolution of the density perturbations. As the linear growth rate $f$ and $f\sigma_8$ are independent of the wave number $k$ in the low redshift regime, we consider the redshifts in the ranges $z=0$ to $z=2$ for our analysis. The figure~\ref{fig:growth} and figure~\ref{fig:sigma} depicts the variation of the linear growth rate $f$ and $f\sigma_8$ as a function of redshift $z$ in the low redshift regime. Since $\zeta$ is negative for the case of model 1, the energy flows from dark matter to dark energy, whereas for the case of model 2, the energy exchange is from dark energy to dark matter. As a result, the dark matter content in the universe will be lower for model 2 than for model 1. This extra amount of dark matter for the case of model 1 influences the matter perturbations, leading to more structure formation in model 1 than in model 2. Thus for model 1, both the linear growth rate $f$ and $f\sigma_8$ will be higher than the $\Lambda$CDM model whereas $f$ and $f\sigma_8$ will be lower for the case of model 2 (see figure~\ref{fig:growth} and figure~\ref{fig:sigma}). In table~\ref{tab:sigma8} we show the values of $\sigma_8$ at redshift $z=0$ for model1, model 2 and the $\Lambda$CDM. We obtained the $\sigma_8$ values using the fiducial values given in table~\ref{tab:model1_table} and table~\ref{tab:model2_table} for model 1 and model 2 respectively. The $\Lambda$CDM model corresponds to $\zeta=0$ and $\omega_x=-1$. Since there is a noticeable difference in the amplitude of  $\sigma_8$ for each model, it will influence the power spectrum of the non-relativistic matter and also the $f\sigma_8$ shown in figure~\ref{fig:sigma}. Furthermore, compared to model 1 and the $\Lambda$CDM, the smaller value of $\sigma_8$ in model 2 also reduces the clustering of galaxies.


As the energy budget of the universe is altered due to the interaction, the interacting dark energy models when compared with $\Lambda$CDM, thus introduce significant changes in various observable probes like CMB power spectrum, matter power spectrum, growth rate and $f\sigma_8$. Since these observable probes are very sensitive to a wide range of cosmological parameters, using them, one can constrain the parameters of the model with great accuracy. Moreover, to detect small coupling constant with great accuracy and to provide tighter constraints on various dark sector parameters ($\zeta, \omega_x $ and $\Omega_ch^2$) we need complementary observations along with CMB data. Different observations probe dark energy and other cosmological parameters in different ways. So we consider cosmic variance limited future CMB experiment PICO along with BAO information from DESI to forecast the constraints on the parameters of the interacting models. One should also use the growth of structure measurements like weak lensing and galaxy clustering information to further constrain the model with greater accuracy. The forecast corresponding to the latest experimental configurations for the growth of structure measurements will be investigated in our future works.



\begin{table}
	\addtolength{\tabcolsep}{7pt}
	\renewcommand{\arraystretch}{1.4}
	\centering
	\begin{tabular}{lcc} 
		\hline 
		\hline
		Parameter   &   Prior   & Prior \\ 
		&    (Model 1)  & (Model 2) \\
		\hline 
		\hline
		{\boldmath$\log(10^{10} A_\mathrm{s})$}
		&  $[1.61, 3.91]$   &  $[1.61, 3.91]$
		\\
		{\boldmath$n_\mathrm{s}   $}
		& 	$[0.8, 1.2]$  &  $[0.8, 1.2]$
		\\
		{\boldmath$100\theta_\mathrm{s}$}
		& $[0.5,10]$  & $[0.5,10]$
		\\
		{\boldmath$\Omega_\mathrm{b} h^2$}
		&  $[0.005, 0.1]$  &  $[0.005, 0.1]$
		\\
		{\boldmath$\Omega_\mathrm{c} h^2$}
		& $[0.001,0.99]$  & $[0.001,0.99]$
		\\    	 
		{\boldmath$\omega_\mathrm{x}  $}
		& $[-0.9999, 0.0]$  & $[-3.000, -1.0001]$
		\\
		{\boldmath$\zeta          $}
		& 	$[-2.0, 0.0]$  & $[0.0, 2.0]$
		\\
		{\boldmath$\tau_\mathrm{reio}$}
		& $[0.01,0.8]$  &  $[0.01,0.8]$
		\\
		\hline 
		\hline
	\end{tabular} 
	\caption{\small The table shows the flat priors imposed on various
		free parameters of model 1 and model 2 for the MCMC analysis. \normalsize
		\label{tab:priors}
	}
\end{table}

\section{Analysis} \label{Sec:Methods}
In this section, we describe the method for the forecast analysis focusing on future generations of CMB and BAO observations. We run Markov Chain Monte Carlo (MCMC) forecasts for the experiment configurations of the PICO~\citep{NASAPICO:2019thw} and DESI~\citep{2013arXiv1308.0847L}, following the commonly used approach described in~\cite{PhysRevD.97.063519},~\cite{PhysRevD.98.083523},~\cite{Delabrouille_2018}, and~\cite{PhysRevD.97.123534}. In this method, we generate mock data set according to some fiducial model. Following the generation of the synthetic data set, one postulates a Gaussian likelihood with a certain experimental noise and applies conventional Bayesian extraction techniques to fit the theoretical predictions for various cosmological models to the mock data. The theoretical predictions of the observational probes are evaluated using the latest version of the Boltzman solver CLASS(V3.2.0)~\citep{Blas_2011,lesgourgues2011cosmic} code. To include the effect of the coupling in the dark sector, we modified the CLASS according to the description detailed in section~\ref{Sec:ide_model}. The parameter values assumed for the generation of the synthetic data are given in table~\ref{tab:model1_table} and table~\ref{tab:model2_table} for model 1 and model 2 respectively. We then perform our forecast analysis using the fiducial values of the parameters that are in agreement with the latest Planck 2018 results~\citep{Planck:2018vyg}. Moreover, since small couplings are preferred by the observational data~\citep{PhysRevD.95.043520,PhysRevD.83.063515, Costa_2017, Yang_2016}, the fiducial values of the coupling constant $\zeta$ and dark energy equation of state parameter $\omega_x$ are considered within $1\sigma$ of the best-fit given in~\cite{Costa_2017}. The posterior distribution for model 1 and model 2 is obtained by sampling the parameter space with an MCMC method. We use the modified version of the MCMC simulator COBAYA~\citep{Torrado_2021} to extract the constraints on the cosmological parameters of both models. The posterior parameter distribution is sampled until the Gelman-Rubin convergence statistic~\citep{Gelman:1992zz} satisfy $R-1 < 0.02$. For the MCMC analysis, we adopt flat priors on the following set of parameters: the baryon and cold dark matter densities $\Omega_bh^2$ and $\Omega_ch^2$, the angular acoustic scale $\theta_s$, the scalar primordial power spectrum amplitude $A_s$, the spectral index $n_s$, the optical depth to reionization $\tau_{reio}$ and finally the free parameters of the model - equation of state of dark energy $\omega_x$ and the coupling constant $\zeta$. The prior ranges for the parameters of model 1 and model 2 are given in table~\ref{tab:priors}.

\begin{table}
	\centering
	\addtolength{\tabcolsep}{0.1mm}
\renewcommand{\arraystretch}{1.2}
	\begin{tabular}{ccccccc}
		\hline
		\hline
		Mission  &  Channel& Beam &$\Delta P$& $\ell_{max}$&$\ell_{min}$& $f_{\rm sky}$\\
		& GHz& arcmin &$\mu K$-& && \\
				& &  &arcmin& && \\
		\hline
		\hline
		PICO  &$75$   & $10.7$ & $4.2$ & $4000$&$2$& $0.75$\\
		&$90$ & $9.5$ & $2.8$ & && \\
		&$108$& $7.9$ & $2.3$ & && \\
		&$129$ & $7.4$ & $2.1$ & && \\
		&$155$  & $6.2$ & $1.8$ & && \\
		&$186$ & $4.3$ & $4.0$ & && \\
		&$223$ & $3.6$ & $4.5$ & && \\
		\hline
		\hline
	\end{tabular}
	\caption{Experimental specifications of PICO.}
	\label{tab:specifications}
\end{table}

\subsection{PICO forecasts} \label{Sec:pico}
We produce forecasts on cosmological parameters of interacting scenarios by using a well-established and robust method widely used in the literature~\citep{PhysRevD.97.063519,PhysRevD.98.083523,Delabrouille_2018,PhysRevD.97.123534}. Under the assumption of fiducial models described previously, we compute the theoretical CMB angular power spectrum for temperature, $C_l^{TT}$, E and B mode polarization,  $C_l^{EE}$ and  $C_l^{BB}$, and cross temperature-polarization,  $C_l^{TE}$, using the modified version Boltzmann code CLASS (V3.2.0). We then consider the experimental noise for the temperature angular spectra of the form~\citep{Perotto_2006},
\begin{equation}
	N_\ell = w^{-1}\exp(\ell(\ell+1)\theta^2/8 \ln 2)~,
\end{equation}  
\begin{figure*}
	\centering 
	\includegraphics[width=1\textwidth]{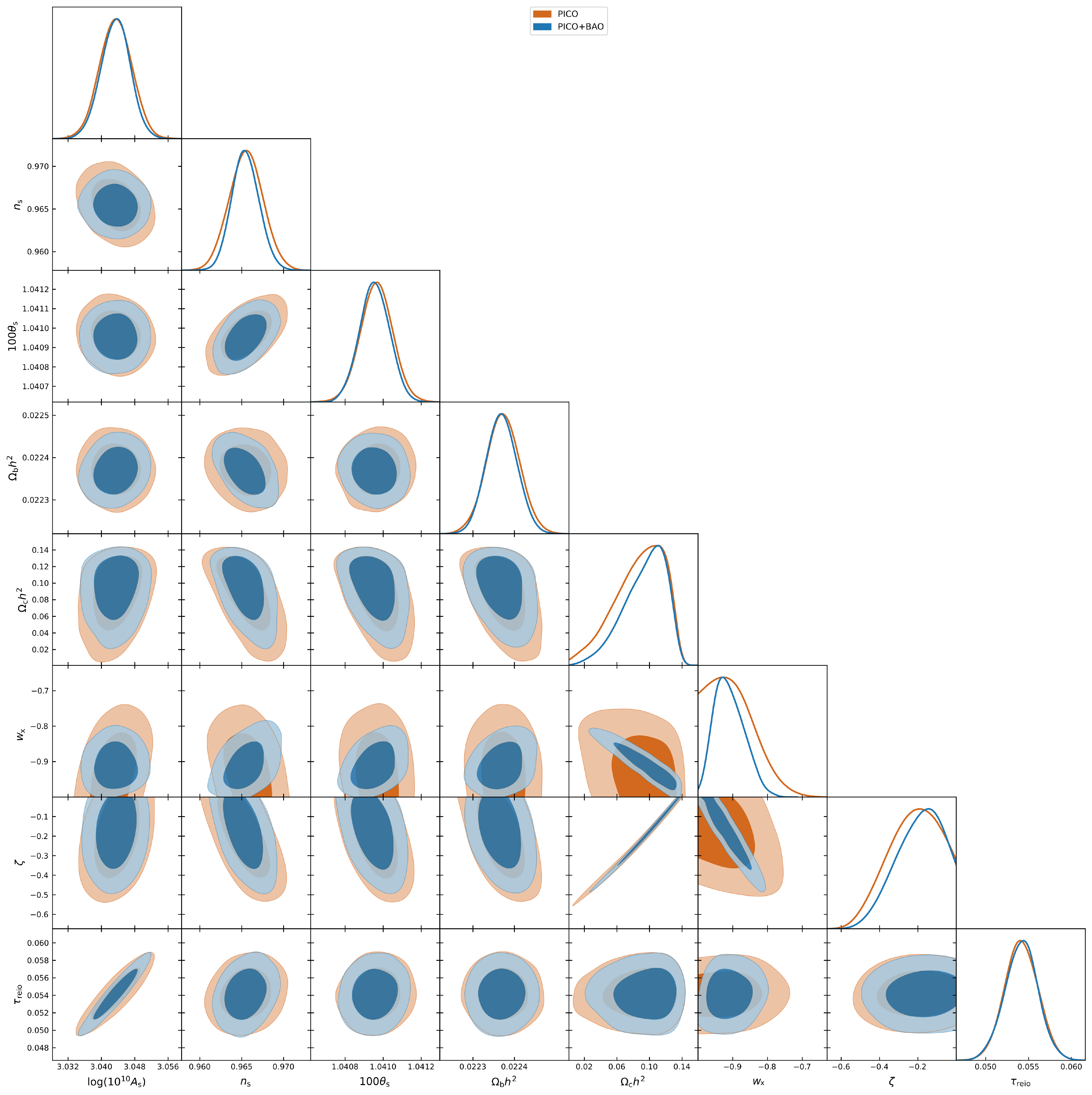}
	\caption{\small Figure showing the posterior distribution of the model parameters for the case of model 1. The contours show $68.3\%$ and $95.5\%$ confidence regions. The quantitative results corresponding to the model 1 are shown in table~\ref{tab:model1_table}.\normalsize
	}
	\label{fig:posteriors_1}
\end{figure*}
where $w^{-1}$ is the experimental sensitivity in units of $\mu$Karcmin, and $\theta$ represents the experimental FWHM angular resolution of the beam. For the case of the noise of polarization spectra, we have $w_p^{-1}= 2 w^{-1}$ (one detector measures two polarization states). We have then produced synthetic realizations of CMB data corresponding to the PICO mission using the experimental specifications of PICO listed in table~\ref{tab:specifications}. Finally, the simulated experimental power spectra are then compared with the theoretical power spectra using a likelihood $\mathcal{L}$ given by,
\begin{equation}
	-2\ln\mathcal{L} = \sum_{\ell} (2\ell + 1)f_{\rm sky}\left(\frac{D_\ell}{|C_\ell|} + \ln\frac{|C_\ell|}{|\hat{C_\ell}|} - 3 \right)~,
\end{equation} 
where $\hat{C} $ and $ C $ represent the theoretical and simulated  power spectra (plus noise), respectively and are defined by :
\begin{align}
	&|C_\ell
	| = C_\ell^{TT}C_\ell^{EE}C_\ell^{BB} -
	\left(C_\ell^{TE}\right)^2C_\ell^{BB}~;  \\
	&|\hat{C}_\ell| = \hat{C}_\ell^{TT}\hat{C}_\ell^{EE}\hat{C}_\ell^{BB} -
	\left(\hat{C}_\ell^{TE}\right)^2\hat{C}_\ell^{BB}~,
\end{align}
while $ D $ is :
\begin{align}
	D_\ell  &=
	\hat{C}_\ell^{TT}C_\ell^{EE}C_\ell^{BB} +
	C_\ell^{TT}\hat{C}_\ell^{EE}C_\ell^{BB} +
	C_\ell^{TT}C_\ell^{EE}\hat{C}_\ell^{BB} \nonumber\\
	&- C_\ell^{TE}\left(C_\ell^{TE}\hat{C}_\ell^{BB} +
	2C_\ell^{TE}C_\ell^{BB} \right)~, \nonumber\\
\end{align}
where $f_{sky}$ is the sky fraction measured by the experiment. The sky fraction $f_{sky}$ and the maximum multipole $\ell$ corresponding to the PICO mission is given in table~\ref{tab:specifications}.

\begin{table*}
	
	\addtolength{\tabcolsep}{16pt}
	\renewcommand{\arraystretch}{1.6}
	\centering
	
	\begin{tabular} { l  c  c  c}
		\hline
		\hline
		Parameter &  Fiducial value  &  PICO  &  PICO + BAO-DESI\\
		\hline
		\hline
		{\boldmath$\log(10^{10} A_\mathrm{s})$} &   $3.044$  &  $3.0435^{+0.0039+0.0076}_{-0.0039-0.0076} $  &  $3.0434^{+0.0035+0.0068}_{-0.0035-0.0070}$\\
		
		{\boldmath$n_\mathrm{s}   $} &    $0.9649$  &   $0.9655^{+0.0020+0.0040}_{-0.0020-0.0038}$  &  $0.9655^{+0.0016+0.0033}_{-0.0016-0.0031}$\\
		
		{\boldmath$100\theta_\mathrm{s}$} &  $1.04092$ &  $1.040965^{+0.000083+0.00017}_{-0.000083-0.00017}$  & $1.040955^{+0.000078+0.00015}_{-0.000078-0.00016}      $\\
		
		{\boldmath$\Omega_\mathrm{b} h^2$} &   $0.02237$ &  $0.022373^{+0.000041+0.000079}_{-0.000041-0.000081}$ & $0.022369^{+0.000037+0.000072}_{-0.000037-0.000070}      $\\
		
		{\boldmath$\Omega_\mathrm{c} h^2$} &  $0.120$  &   $0.088^{+0.039+0.048}_{-0.021-0.060}$  &  $0.094^{+0.032+0.042}_{-0.018-0.053}   $\\
		
		{\boldmath$\omega_\mathrm{x}  $} &  $-0.95$ & $ -0.900^{+0.029}_{-0.098}\,,<-0.790 $ &  $-0.909^{+0.035+0.084}_{-0.052-0.077}  $\\
		
		{\boldmath$\zeta          $} & $-0.08$ &   $-0.223^{+0.18}_{-0.096}\,,>-0.451   $  & $-0.198^{+0.16}_{-0.083}\,, >-0.411  $\\
		
		{\boldmath$\tau_\mathrm{reio}$} & $0.0544$ & $0.0542^{+0.0020+0.0038}_{-0.0020-0.0039}  $ & $0.0542^{+0.0019+0.0038}_{-0.0019-0.0039}          $\\
		\hline
		\hline
	\end{tabular}
	\caption{\small Forecasted constraints at 68\% and 95\% confidence levels from future CMB experiment PICO and BAO information the DESI galaxy survey for the case of model 1. The fiducial model consists of 8 parameters. We obtain the fiducial parameter values within $1\sigma$ confidence levels for the cases PICO and PICO+BAO(DESI). The parameter constraints are improved when BAO information is integrated with CMB information.
		\normalsize
		\label{tab:model1_table}
	}
\end{table*}

\subsection{DESI (BAO) forecasts} \label{Sec:desi}
The acoustic waves triggered by small overdensities in the very early evolutionary stages of the Universe left an imprint on the distribution of matter when the plasma turned into neutral atoms. This imprinted pattern can be measured as a function of redshift using the BAO technique.
These BAO observations will allow us to probe diverse aspects of cosmology like signatures of inflation,  neutrino mass hierarchy, the nature of dark energy, and thus provide tighter constraints on several cosmological parameters~\citep{2013arXiv1308.0847L}.

For the future BAO dataset, we consider the Dark Energy Spectroscopic Instrument (DESI) experiment~\citep{2013arXiv1308.0847L}. To simulate BAO observations, we use the volume-averaged distance $D_V$ defined as,
\begin{equation}
	D_V(z)\equiv \left(\frac{(1+z)^2D_A(z)^2cz}{H(z)}\right)^\frac{1}{3}~,
\end{equation}
where $H(z)$ is the Hubble parameter and $D_A$ is the angular diameter distance. The angular diameter distance $D_A$ is given by,
\begin{equation}
	D_A(z) = \frac{c}{1+z}\int_0^z\frac{dz}{H(z)}.
\end{equation}
Assuming the fiducial model described in table~\ref{tab:model1_table} and table~\ref{tab:model2_table}, we compute the theoretical values of the ratio $r_s/D_V$. Here $r_s$ represents the sound horizon at the drag epoch when photons and baryons decouple. The theoretical values of the ratio $r_s/D_V$ are computed in the redshift range $z = [0.15 - 1.85]$. Moreover, the sound horizon $r_s$ is defined as,
\begin{equation}
	r_s = \int_{z_d}^\infty \frac{c_s(z)}{H(z)}dz,
\end{equation}
where $z_d$ is the redshift at the drag epoch and $c_s = \frac{c}{\sqrt{3(1+R)}}$ the sound speed, which depends on the ratio of baryon to photon density, with $R=3\rho_b/4\rho_{\gamma}$. 
We then compare the simulated BAO datasets with the theoretical values of $r_s/D_V$ through a Gaussian prior. Though it is possible to forecast BAO data considering $D_A/r_s$ and $H(z)$ as independent measurements, there exists a small tension ($\sim$ 1$\sigma$) between the current constraints from $D_A/r_s$ and $H(z)$~\citep{Addison_2018}. Due to the difficulty of properly accounting for this small tension between $D_A/r_s$ and $H(z)$, we follow the approaches in~\cite{Addison_2018} and ~\cite{PhysRevD.98.083523} to generate the datasets (see table 2 in~\cite{PhysRevD.98.083523}) for the DESI experiment.

\section{Results}\label{sec: results}

\begin{table}
	\addtolength{\tabcolsep}{10pt}
	\renewcommand{\arraystretch}{1.6}
	\centering
	\begin{tabular}{l c}
		\hline 
		\hline
		\text{Parameter} &  \text{Planck 2018}    \\
		
		\hline
		\hline
		{\boldmath$\log(10^{10} A_\mathrm{s})$}      & $3.044 \pm 0.014$         \\
		{\boldmath$n_\mathrm{s}   $}       & $0.9649 \pm 0.0044$       \\
		{\boldmath$100\theta_\mathrm{s}$}       & $1.04092 \pm 0.00031$     \\
		{\boldmath$\Omega_\mathrm{b} h^2$}      & $0.02236 \pm 0.00015$     \\
		{\boldmath$\Omega_\mathrm{c} h^2$}  & $0.1201 \pm 0.0014$       \\
		{\boldmath$\tau_\mathrm{reio}$}       & $0.0544 \pm 0.0073 $      \\
		\hline 
		\hline
	\end{tabular}
	\caption{\small The mean values with $68\%$ intervals for the parameters of the $\Lambda$CDM model from the latest Planck 2018 results~\citep{Planck:2018vyg}. There is a significant improvement in the constraints for all the parameters (excluding $\Omega_ch^2$) when CMB information from future generation PICO experiment is taken into account (see table~\ref{tab:model1_table} and table~\ref{tab:model2_table}). \normalsize}
	\label{tab:best_fit_planck}
\end{table}

\begin{table}
	
	\addtolength{\tabcolsep}{1pt}
	\renewcommand{\arraystretch}{1.6}
	\centering
	
	\begin{tabular} { l  c  c  c}
		\hline
		\hline
		Parameter &  Model 1  &  Model 2  \\
		&  \text{\small (Planck 2018+BAO)} & (\small Planck 2018+BAO) \\		
		\hline
		\hline
		{\boldmath$\log(10^{10} A_\mathrm{s})$} &   $3.045^{+0.034}_{-0.032}   $  &  $3.045^{+0.032}_{-0.030}   $  \\
		
		{\boldmath$n_\mathrm{s}   $} &    $0.9658^{+0.0082}_{-0.0084}   $  &   $0.9657^{+0.0081}_{-0.0082}   $  \\
		
		{\boldmath$100\theta_\mathrm{s}$} &  $1.0438^{+0.0043}_{-0.0031}   $ &  $1.0401^{+0.0010}_{-0.0010}   $  \\
		
		{\boldmath$\Omega_\mathrm{b} h^2$} &   $0.02239^{+0.00029}_{-0.00027}   $ &  $0.02239^{+0.00029}_{-0.00028}   $ \\
		
		{\boldmath$\Omega_\mathrm{c} h^2$} & $0.076^{+0.046}_{-0.058}$  &   $0.134^{+0.014}_{-0.014}$  \\
		
		{\boldmath$\omega_\mathrm{x}  $} &  $<-0.80$ & $-1.094^{+0.094}_{-0.099}$ \\
		
		{\boldmath$\zeta  $} &  $>- 0.24$ & $< 0.101$ \\
		
		{\boldmath$\tau_\mathrm{reio}$} &$0.055^{+0.017}_{-0.015}$ & $0.055^{+0.016}_{-0.015}$ \\
		\hline
		\hline
	\end{tabular}
	\caption{\small  The mean values with $95\%$ confidence intervals obtained in~\citep{PhysRevD.101.083509} for the parameters of model 1 and model 2. Here the constraints are derived by integrating  the latest Planck 2018 data  with BAO information from 6dFGS, SDSS-MGS and BOSS DR12 surveys. (note: here the coupling constant $\zeta$ is the negative of the coupling constant considered in~\citep{PhysRevD.101.083509} to maintain same sign convention in both works.)
		\normalsize
		\label{tab:earlier}
	}
\end{table}


Following the methodology described in section~\ref{Sec:Methods}, we compute the expected constraints on the parameters of model 1 and model 2 from future generation CMB experiment PICO and BAO information from the DESI galaxy survey. Using a modified version of COBAYA and a convergence diagnostic based on Gelman and Rubin statistics, we carry out MCMC analysis. We use the GetDist~\citep{2019arXiv191013970L} software package to statistically analyse the MCMC findings.

\begin{figure*}
	\centering 
	\includegraphics[width=1\textwidth]{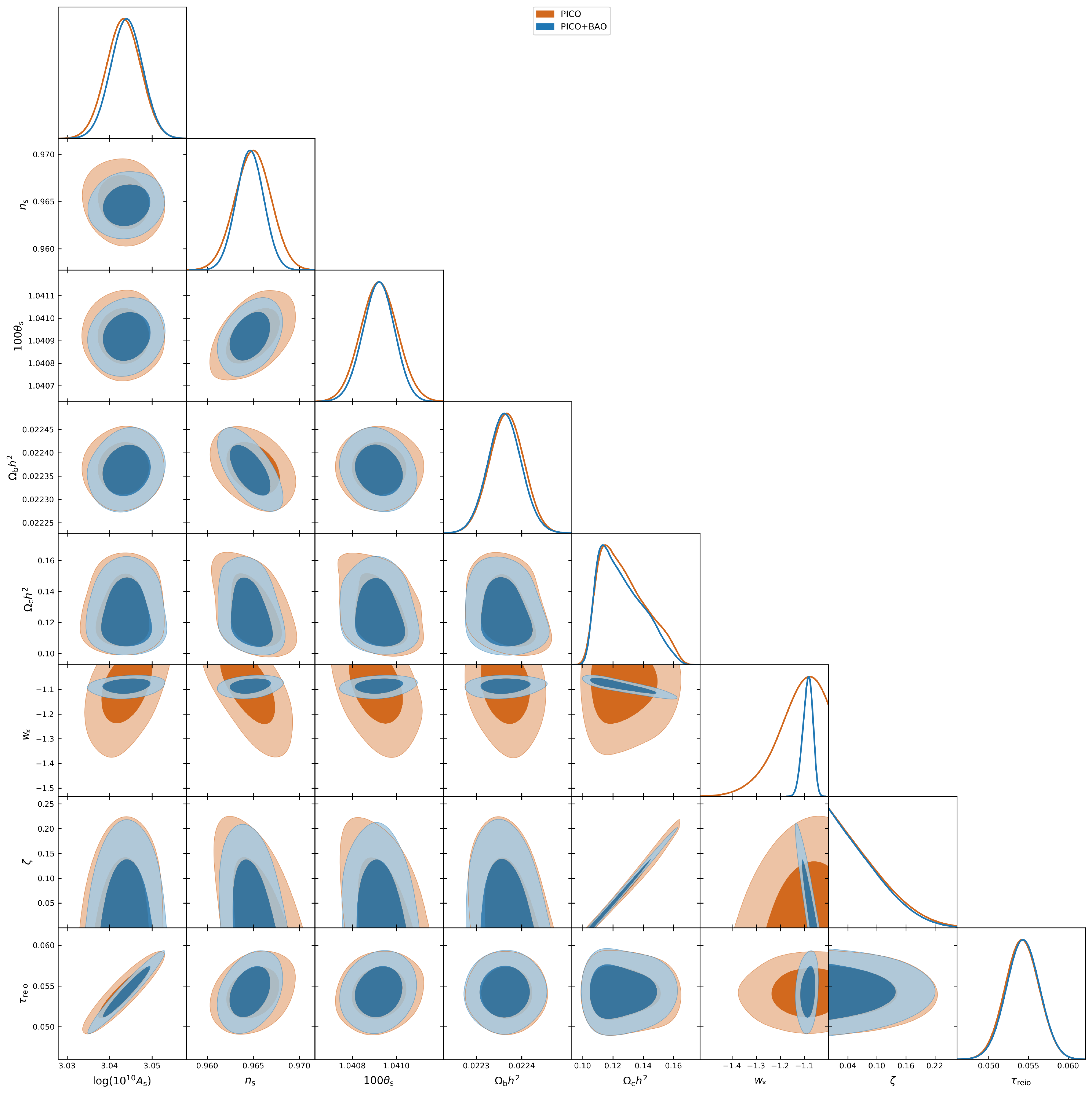}
	\caption{\small Figure showing the posterior distribution of the model parameters for the case of model 2. The contours show $68.3\%$ and $95.5\%$ confidence regions. The quantitative results corresponding to the model 2 are shown in table~\ref{tab:model2_table}. \normalsize
	}
	\label{fig:posteriors_2}
\end{figure*}

Since the coupling constant $\zeta$ is negative for the case of model 1, the energy exchange is from dark matter to dark energy, and the equation of state parameter of dark energy is in the quintessence regime. In section~\ref{Sec:effects_on_probes}, we investigated the effect of the coupling constant $\zeta$ and dark energy equation of state parameter $\omega_x$ on CMB temperature and matter power spectrum. The low-$\ell$ modes of the CMB power spectrum are influenced by the parameters $\omega_x$ and $\zeta$. Since the cosmic variance is greater in the lower multipole areas, it is challenging to discern between the influence of the parameters $\zeta$ and $\omega_x$. So the constraining power of the CMB experiment will be limited for these parameters. Figure~\ref{fig:posteriors_1} displays the 1-d marginalized posterior distributions and 2-d marginalized constraint contours for the parameters of model 1. The contours display the $1\sigma$ region of 68 percent confidence and the $2\sigma$ region of 95 percent confidence, with the darker colour denoting the more likely outcomes. The quantitative findings with $68\%$ and $95\%$ confidence levels from the CMB experiment PICO and BAO information from DESI galaxy survey for the case of model 1 are shown in table~\ref{tab:model1_table}. It is interesting to note that there is a significant improvement in the parameter constraints compared to the constraints from the latest Planck data (see table~\ref{tab:best_fit_planck}). Moreover, due to the interaction between the dark sectors, the parameters $\Omega_ch^2$, $\omega_x$ and $\zeta$ are weakly constrained. But when the BAO information from DESI galaxy survey is integrated with PICO experiment, there is a notable improvement in the parameter constraints, especially for the dark sector parameters $\Omega_ch^2$, $\omega_x$ and $\zeta$ ( see table~\ref{tab:model1_table} and table~\ref{tab:best_fit_planck}). Since the effect of $\omega_x$ is only at cosmic variance dominated low-$\ell$ modes of the CMB power spectrum, the addition of BAO data can significantly improve its constraints compared to CMB data alone. It is also worth mentioning that, as the constraints become tighter with the addition of BAO, the correlation between the parameter pairs ($\omega_x$, $\Omega_ch^2$) and ($\omega_x$, $\zeta$) are clearly visible. We also obtain the fiducial values of the parameters within $1\sigma$ standard deviations for the cases PICO and PICO+DESI. Moreover, the accuracy of the parameters of model 1 has increased significantly when compared with the most recent real data constraints in~\cite{PhysRevD.101.083509} derived by using Planck 2018 data and BAO information from 6dFGS, SDSS-MGS, and BOSS DR12 surveys. The $95\%$ confidence level constraints obtained  in~\cite{PhysRevD.101.083509} by using Planck 2018 data and BAO are shown in table~\ref{tab:earlier}. So by comparing table~\ref{tab:model1_table} and table~\ref{tab:earlier}, we can see that the accuracy of the parameters $\log(10^{10} A_\mathrm{s}), n_\mathrm{s}, 100\theta_\mathrm{s}, \Omega_\mathrm{b} h^2, \,$and$ \,\tau_\mathrm{reio}$ are increased by  5, 2.48, 28.6, 4.02 and 4.47 times respectively when PICO and BAO from DESI is incorporated instead of  Planck and BAO for the case interactiong dark energy model 1. Moreover the constraints on the dark sector parameters ($\Omega_\mathrm{c} h^2, \omega_x$ and $\zeta$) also reflects the degeneracies between them~\citep{PhysRevD.83.063515}. As a result, an increase in the accuracy of parameters ($\Omega_\mathrm{c} h^2 \,\,$and$\,\, \omega_x$ in model 1) decreases the accuracy of the other dark sector parameter ($\zeta$ in model 1). It is also worth noting that the equation of state parameter $\omega_x$ of model 1, which was unbounded with joint constraints of Planck 2018 and BAO data, is well constrained when PICO is integrated with DESI. Therefore, the future generation CMB experiment PICO and DESI galaxy survey can constrain cosmological parameters with excellent accuracy.

The posterior distribution for the parameters for the case of model 2 is shown in figure~\ref{fig:posteriors_2}, and the corresponding quantitative results are summarized in table~\ref{tab:model2_table}. Since $\zeta$ is positive for model 2, the energy flow is from dark energy to dark matter, and the equation state parameter of dark energy is in the phantom regime. Even when the dark energy equation state is in the phantom regime, the effect of $\zeta$ and $\omega_x$ is mainly at the cosmic variance dominated low-$\ell$ modes of the CMB power spectrum. So it is challenging for CMB experiments to detect low coupling constant $\zeta$ and dark energy equation of state parameter $\omega_x$. As a result, using the CMB data alone, the dark sector parameters $\Omega_ch^2$, $\omega_x$ and $\zeta$ are weakly constrained. But it is also noteworthy that there is a significant improvement in the constraints of the other parameters compared to the latest Planck results~\citep{Planck:2018vyg} based on the $\Lambda$CDM model(see table~\ref{tab:best_fit_planck}). Moreover, there are noticeable improvements in the parameter constraints when the BAO information is integrated with CMB data, especially for the parameter $\omega_x$. As the effect of $\omega_x$ is only at cosmic variance dominated low-$\ell$ modes of the CMB power spectrum, the constraints on $\omega_x$ are unbounded when we use CMB data alone. But with the integration of BAO data with CMB, one can obtain tighter constraints for $\omega_x$ (see table~\ref{tab:model2_table}). We also obtain the fiducial values of the parameter of model 2 within $1\sigma$ confidence levels for the cases PICO and PICO+BAO(DESI). Compared to the constraints on the dark energy equation of state parameter $\omega_x$ in the quintessence regime (see table~\ref{tab:model1_table}), the constraints on $\omega_x$ are tighter when it is in the phantom regime (see table~\ref{tab:model2_table}). Moreover, it is also interesting to note that, as the constraints become tighter with the addition of BAO, the correlation between the parameter pairs ($\omega_x$, $\Omega_ch^2$) and ($\omega_x$, $\zeta$) are clearly visible. Additionally, when PICO is combined with BAO from the DESI galaxy survey, the constraints on the parameters of model 2 are much improved as compared to earlier study in~\cite{PhysRevD.101.083509}(see table~\ref{tab:earlier}). It is evident from table~\ref{tab:earlier} and table~\ref{tab:model2_table} that the constraints on the parameters $\log(10^{10} A_\mathrm{s}), n_\mathrm{s}, 100\theta_\mathrm{s}, \Omega_\mathrm{b} h^2, \,$and$ \,\tau_\mathrm{reio}$ are increased by  4.63, 3, 7.69, 3.97 and 4.10 times respectively for the case of PICO+DESI. Similarly, as seen previously for the case of model 1, an increase in the accuracy of one dark sector parameter ($\omega_x$) of model 2 decreases the accuracy of the other dark sector parameters owing to the degeneracies between them~\citep{PhysRevD.83.063515}. Thus, to detect small coupling and further constrain all dark sector parameters, we need more complementary observations like the growth of structure measurements like weak lensing and galaxy clustering information, which will be discussed in a future article.

\begin{table*}
	
	\addtolength{\tabcolsep}{16pt}
	\renewcommand{\arraystretch}{1.6}
	\centering
	
	\begin{tabular} { l  c  c  c}
		\hline
		\hline
		Parameter &  Fiducial value  &  PICO  &  PICO + BAO-DESI\\
		\hline
		\hline
		{\boldmath$\log(10^{10} A_\mathrm{s})$} &   $3.044$  &   $3.0432^{+0.0038+0.0073}_{-0.0038-0.0076}$  &  $3.0440^{+0.0035+0.0069}_{-0.0035-0.0070}   $\\
		
		{\boldmath$n_\mathrm{s}   $} &    $0.9649$  &    $0.9650^{+0.0018+0.0036}_{-0.0018-0.0036}$  & $0.9646^{+0.0014+0.0027}_{-0.0014-0.0028}$\\
		
		{\boldmath$100\theta_\mathrm{s}$} &  $1.04092$ &  $1.040923^{+0.000079+0.00016}_{-0.000079-0.00015}$  & $1.040919^{+0.000068+0.00013}_{-0.000068-0.00014}     $\\
		
		{\boldmath$\Omega_\mathrm{b} h^2$} &   $0.02237$ &  $0.022367^{+0.000036+0.000069}_{-0.000036-0.000071}$ & $0.022363^{+0.000036+0.000073}_{-0.000036-0.000070}  $\\
		
		{\boldmath$\Omega_\mathrm{c} h^2$} &  $0.120$  &   $0.1267^{+0.0078+0.029}_{-0.020-0.022}$  &  $0.1255^{+0.0076+0.027}_{-0.019-0.021}   $\\
		
		{\boldmath$\omega_\mathrm{x}  $} &  $-1.08$ & $ -1.13^{+0.12}_{-0.042}\,,> -1.30 $ &  $-1.086^{+0.023+0.034}_{-0.016-0.039}  $\\
		
		{\boldmath$\zeta          $} & $0.05$ &  $< 0.0922 \,, <0.178    $  & $< 0.0910\,, <0.165   $\\
		
		{\boldmath$\tau_\mathrm{reio}$} & $0.0544$ & $0.0542^{+0.0020+0.0040}_{-0.0020-0.0040}    $ & $0.0543^{+0.0020+0.0039}_{-0.0020-0.0039}      $\\
		\hline
		\hline
	\end{tabular}
	\caption{\small Forecasted constraints at 68\% and 95\% confidence levels from future generation CMB experiment PICO and information from the BAO DESI galaxy survey for the case of model 2. The fiducial model consists of 8 parameters. We obtain the fiducial parameter values within $1\sigma$ confidence levels for the cases PICO and PICO+BAO(DESI). The constraints on the parameters are tighter when BAO information is integrated with CMB information, especially for the case of equation of state parameter of dark energy.
		\normalsize
		\label{tab:model2_table}
	}
\end{table*}

\section{Conclusions}\label{sec: conclusion}

The cosmological models with energy exchange between the dark sectors have attracted much attention over the years since they can give rise to comparable energy densities at the present epoch. In this study, we focused our interest on phenomenological interacting dark energy models and examined the effects of two observational probes on the parameter constraints, especially the CMB temperature and polarisation spectrum and BAO data from a redshift range of $0.15\leq z \leq 1.85$. With the advent of precision cosmology, future generations of cosmological probes are expected to provide very accurate observational data to measure cosmological parameters with unprecedented precision. So we exploited the capabilities of high precision cosmic variance limited future CMB polarization experiment PICO along with BAO information from the DESI experiment to constrain the parameters of the interacting dark sector and distinguish them from models in which there is no interaction between the dark sector. In our analysis, we use the MCMC exploration of the parameter space as it does not make any assumptions about the Gaussianity or otherwise of the parameter probabilities and is therefore expected to give more reliable results than the traditional Fisher forecast.  It also has the added advantage that one can use the same parameter extraction pipeline when the real data becomes available.

Based on the stability of the cosmological perturbations, we considered two possibilities for the interaction scenario, i.e., model 1 ($\zeta<0, -1/3>\omega_x>-1$) and model 2 ($\zeta>0, \omega_x<-1$). We then investigated the effect of these interacting models on different observable probes and distinguished them from models in which there is no energy exchange between dark sectors. For model 1, since the energy transfer is from dark matter to dark energy, the dark matter content in the past is higher compared to the $\Lambda$CDM model. Having more dark matter content in the past leads to an overall suppression in the CMB angular power spectrum. An increase in the matter content of the universe in model 1 due to the presence of interaction drives the matter power spectrum up and the CMB spectrum down. This effect is exactly the opposite for the case of model 2, where the coupling constant is positive, and the dark energy equation of state parameter is in the phantom regime. The effect of the dark energy equation of state parameter $\omega_x$ and coupling constant $\zeta$ are mainly on the cosmic variance dominated low-$\ell$ modes of the CMB power spectrum. Moreover, the variations in the dark sector budget in both models also affect the growth of the large-scale structures compared to the $\Lambda$CDM model. The extra amount of dark matter in model 1 influences the matter perturbations and leads to more structure formation. As a result, the linear growth rate and galactic clustering are highest for model 1 when compared to $\Lambda$CDM and model 2.

By comparing the interacting dark energy models against high-sensitive future generation CMB experiment PICO and BAO information from the DESI galaxy survey, we forecast the parameter constraints for both models in section~\ref{sec: results}. In comparison with the latest Planck 2018 results~\citep{Planck:2018vyg}, there is a significant improvement in the parameter constraints when simulated CMB data from the PICO mission is used for the forecast analysis. As the constraints on the parameters are tighter, the cosmological parameters are measured with unprecedented precision. Moreover, when we introduce interaction between the dark sectors, parameters $\Omega_ch^2$, $\zeta$ and $\omega_x$ are weakly constrained. But when the BAO information from the DESI galaxy survey is integrated with the PICO experiment, there is a notable improvement in the parameter constraints for both models, especially for these dark sector parameters. Since the effect of $\omega_x$ and $\zeta$ is mainly at cosmic variance dominated low-$\ell$ modes of the CMB power spectrum, the addition of BAO data can significantly improve its constraints compared to CMB data alone.

We obtain the fiducial values of the parameters of the two models within $1\sigma$ standard deviation error. Our current analysis also demonstrates a significant increase in the accuracy of the parameters of both interacting dark energy models compared to the latest real data constraints obtained in~\cite{PhysRevD.101.083509}. Additionally, we have shown that, for model 1, the coupling constant $\zeta$ can be constrained at the $1\sigma$ confidence level. However, for model 2, i.e. when the energy exchange is from dark energy to dark matter, we can only obtain the upper bound on $\zeta$. Thus to detect small coupling and to further constrain parameters of models with a positive coupling constant, we need more complementary observations like cosmic chronometers~\citep{PhysRevD.94.023508,Moresco_2016}, 21-cm cosmology~\citep{Battye:2016qhf,Bull_2015}, redshift dependence of the Alcock-Paczynski effect~\citep{Li_2016} and gravitational waves~\citep{Caprini_2016}. These observational probes may even be able to further improve the constraints on the interacting dark sector models and measure the parameters with unprecedented precision.
\\

\section*{Acknowledgements}
For the computations required by this work, we utilise the Kanad high-performance computing facility of IISER Bhopal.

\section{Data Availability}
The data underlying this article will be shared on reasonable request to the corresponding author.



\bibliographystyle{mnras}
\bibliography{ms} 



\bsp	
\label{lastpage}
\end{document}